\documentclass[prd,aps,11pt,superscriptaddress]{revtex4-1}
\usepackage[utf8]{inputenc}
\usepackage[dvipsnames]{xcolor}
\usepackage{graphicx}
\graphicspath{ {./figures/} }
\usepackage{textcomp}

\begin{document}

\title{A very high energy hadron collider on the Moon}

\author{James Beacham}
\email{j.beacham@cern.ch}
\thanks{ORCID: \href{https://orcid.org/0000-0003-3623-3335}{0000-0003-3623-3335}}
\affiliation{Duke University, Durham, N.C., United States}

\author{Frank Zimmermann}
\email{frank.zimmermann@cern.ch}
\affiliation{CERN, Meyrin, Switzerland}

\date{\protect\today}

\begin{abstract}
The long-term prospect of building 
a hadron collider around the circumference of a great circle of the Moon is sketched. A Circular Collider on the Moon (CCM) of $\sim$11000~km in circumference could reach a proton-proton center-of-mass collision energy of 14~PeV --- a thousand times higher than the Large Hadron Collider at CERN --- optimistically assuming a dipole magnetic field of 20~T. Several aspects of such a project are presented, including siting, construction, availability of necessary materials on the Moon, and powering, as well as a discussion of future studies and further information needed to determine the more concrete feasibility of each. Machine parameters and vacuum requirements are explored, and an injection scheme is delineated. Other unknowns are set down. Due to the strong interest from multiple organizations in establishing a permanent Moon presence, a CCM could be the (next-to-) next-to-next-generation discovery machine for high-energy particle physics and a natural successor to next-generation machines, such as the proposed Future Circular Collider at CERN or a Super Proton-Proton Collider in China, and other future machines, such as a Collider in the Sea, in the Gulf of Mexico. A CCM would serve as an important stepping stone towards a Planck-scale collider sited in our Solar System.  
\end{abstract}

\maketitle

\newpage
\section{Introduction}\label{sec:intro}

After the discovery of the Higgs boson by the ATLAS and CMS collaborations at the Large Hadron Collider (LHC) at CERN in 2012~\cite{Aad:2012tfa,Chatrchyan:2012ufa}, high-energy particle physics finds itself with two main goals for the future. The first is to accomplish a high-precision study of 
the Higgs and other Standard Model (SM) particles and parameters. The second is to attain higher center-of-mass collision energies with hadrons, to explore unexplored parameter space and open up the prospect of new particle discoveries. Both of these goals are the foci of the proposed next-generation circular collider projects such as the Future Circular Collider (FCC)~\cite{Abada:2019lih,Abada:2019zxq,Benedikt:2018csr} at CERN; the Circular Electron-Positron Collider (CEPC) and its potential successor, the Super Proton-Proton Collider (SPPC)~\cite{CEPC-SPPCStudyGroup:2015csa,CEPC-SPPCStudyGroup:2015esa,CEPCStudyGroup:2018rmc,CEPCStudyGroup:2018ghi} in China; or, more exotically, a Collider in the Sea (CitS) floating in the Gulf of Mexico, as has been proposed by a few U.S.-based collider experts~\cite{McIntyre:2017ibd}.
The FCC and the CEPC/SPPC projects could potentially be realized sometime in the next few decades, as both initiatives have extensive interest from the global community, as well as advanced conceptual design reports (CDRs). The CitS idea could be potentially envisioned on a longer time-frame, having relatively recently been conceptualized. The existence of such ideas, whether in the form of more mature CDRs or as conceptual outlines, indicates the strong interest by the particle physics community in reaching higher collision energies.
During eventual proton-proton collisions, these machines could reach center-of-mass energies of 80--120 TeV (or 500 TeV in the case of the CitS), a substantial increase over the current energy of 13 TeV at the LHC.

The state of particle physics after the Higgs discovery, however, is such that there are multiple unsolved mysteries of physics, but few (or no) hints as to the mass scale of new particles or phenomena, that, if discovered, could help solve these. 
The SM of particle physics, for all its successes, is known to be incomplete. For example, it does not contain gravity, does not contain dark matter, cannot explain dark energy, doesn't account for the matter / antimatter asymmetry of the universe, doesn't account for the values of neutrino masses, and offers no comprehensive explanation for why its structure is the way it is. There are myriad excellent theoretical models or classes of models that provide extensions to the SM to (at least partially) account for these limitations and these models predict a wide variety of new particles and/or forces to be discovered, at higher masses than are directly accessible at current colliders or with small enough cross sections as to have eluded existing datasets. Additionally, the SM itself makes a large number of predictions, all of which, thus far, have been confirmed, but many of which are either currently being tested with greater precision (such as the magnetic moment of the muon~\cite{Muong-2:2021vma}) or can only be tested with larger datasets in the future, e.g., at the High-Luminosity LHC~\cite{2015hllh.book.....B} or future machines like the FCC.

However, the existence of specific theoretical models is not the reason higher-energy colliders should be built. The broader question to be addressed by the particle physics community is as follows: At what energy does the validity of the SM break down, if at all? In principle the SM could be valid all the way to the Planck energy, $10^{16}$ TeV. Thus, a full and complete understanding of the SM---and an identification of the point at which it becomes invalid, thus providing a long-sought window toward new physics to explain the open questions---will not be possible without a series of collider experiments that test the SM at successively higher energies and simultaneously explore the new parameter space searching for deviations from background expectation or generic anomalies.
New particles could appear at any energy up to the Planck energy, and, as such, the full range of energies up to and including the Planck scale must be tested. As a result of this, any future collider experiment already has ample justification for its physics program. The future of collider physics is exploratory and experimental, irrespective of specific theory motivation, because the motivations are simple: There are a large number of open questions to answer, and new particles to answer them could arise at any energy. The physics cases for an FCC, a CEPC/SPPC, and a CitS are clear, but stretching beyond those projects requires exploratory experiments that reach as high a center-of-mass collision energy as possible.

\begin{figure}[h]
    \centering
    \includegraphics[width=0.9\textwidth]{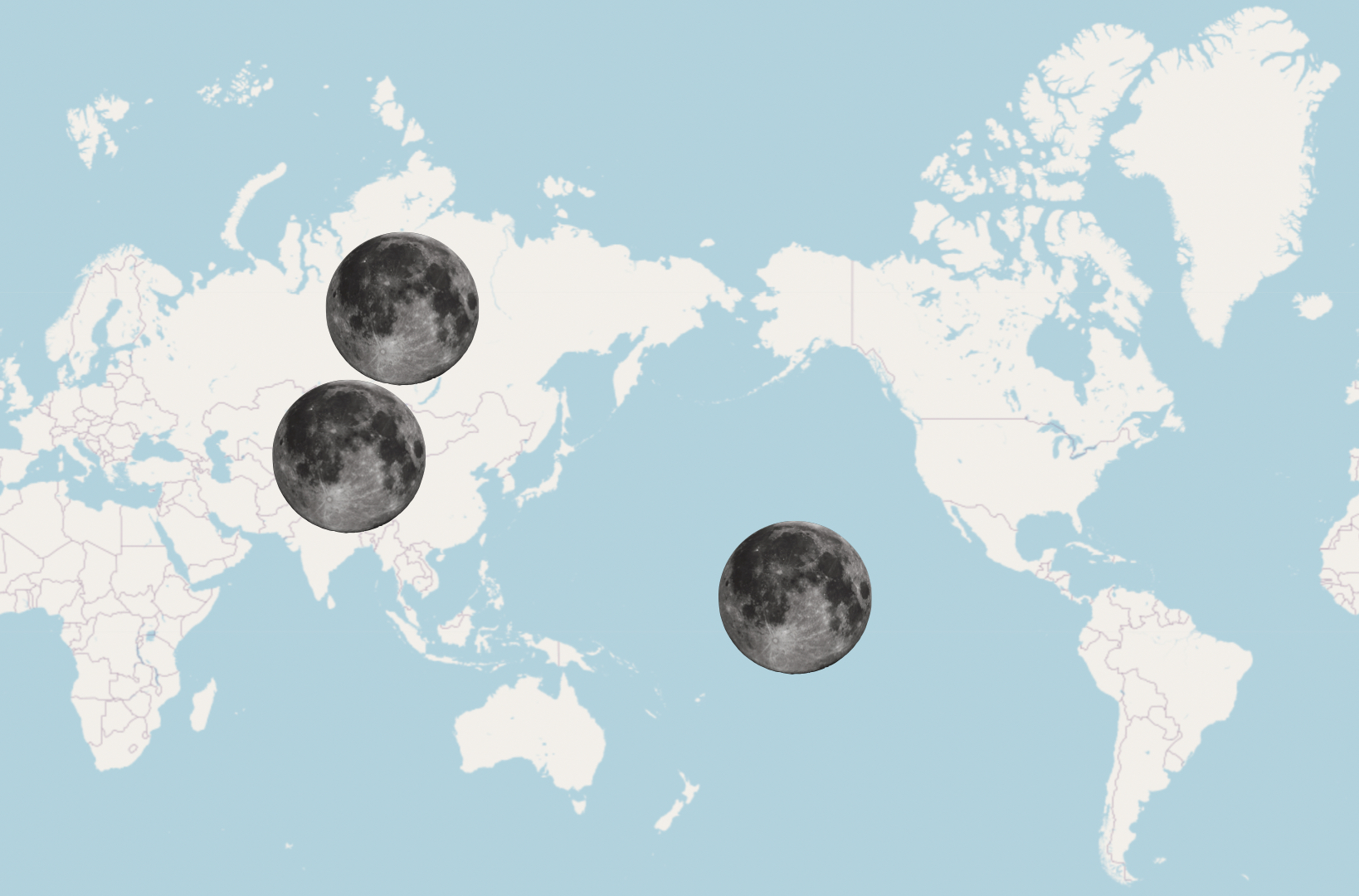}
    \caption{Three potential Earth-based sites for a circular collider approximately the same size as a collider encircling the Moon of $\sim$11000 km in circumference, represented by images of the Moon overlaid on a map of the surface of the Earth. Each potential Earth-based site for such a large collider project is accompanied by significant geographical, technological, or political challenges. Adapted from Ref.~\cite{openstreetmap} and Ref.~\cite{wikimoon}.}
    \label{fig:CCMOnEarth}
\end{figure}

Building a hadron collider much larger than those proposed for the next generation (of $\sim$80--100 or 1900 km in circumference) on Earth will likely prove politically and geographically challenging to a prohibitive degree. While multiple potential circular tunnel trajectories on the order of 10,000 km in length can at least naively be identified on Earth, they each are subject to significant drawbacks. (See Figure~\ref{fig:CCMOnEarth}.)

For example, the longer the circular tunnel, the more likely that multiple access shafts will be located in populated areas on the surface that may displace local residents or disrupt the local ecology. Additionally, for potential sites on land — e.g., located entirely within Asia, traveling under Russia, China, Mongolia, Kazakhstan, India, etc. — a 10,000-km tunnel would likely traverse multiple types of rock, soil, and underground water sources and would pass under large mountain ranges (e.g., the Himalayas), leading to potentially insurmountable tunnel boring challenges, as well as a high probability of disrupting water sources or other resources used by humans. The challenges with siting a 10,000-km circular collider entirely within the Pacific or Atlantic Oceans — in contrast to the CitS, which utilizes a likely achievable neutral buoyancy in the Gulf of Mexico — should be clear.

Moreover, longer circular colliders on Earth lead to increasingly difficult technological and geological challenges to ensure the stability of the experiment. Creating and maintaining a 10,000-km-long vacuum for a proton beam, while entirely possible, will be costly on Earth, whereas the vacuum on and under the surface of the Moon is free. Additionally, as the Earth is highly geologically active, a longer collider is more susceptible to the risk of earthquakes and volcanic eruptions.

As a result, while a 10,000-km-scale circular collider on Earth is of course technically possible, it carries with it a substantial number of prominent issues that would make it less attractive compared to other potential basic science projects here on Earth.

In contrast, the Moon presents a potentially ideal site for the next great step forward in collider physics, primarily due to the fact that a return by humans to the Moon has a high probability of being realized in the next century. Both NASA and private companies are interested in establishing a permanent presence on the Moon, and the U.S. Department of Energy has encouraged the international particle physics community to collaborate in this endeavor~\cite{siegrist}. 
This consideration evinces the possibility of a Moon-based PeV-scale collider. 

Indeed, through its Artemis program \cite{artemis}, launched in 2017, NASA aims
at establishing a 
sustainable presence on the Moon by 2028. 
In preparation for this, Moon deliveries are already starting as early as 2021, in the frame of the 
Commercial Lunar Payload Services (CLPS) initiative \cite{CPLS}.   
The near-term goal is to explore the entire surface of the Moon with human and robotic explorers. 
This exploration will provide relevant information for 
optimizing the placements of future large-scale lunar  infrastructures.  
Potential materials and structural concepts for second-generation habitat construction on the Moon are 
also rather well established already. 
For example, Ref.~\cite{doi:10.1061/0893} discusses a simple structural frame for erection and concludes that a lunar habitat can be built with a reasonable factor of safety and using existing technology. 

We note that the United Nations 
Outer Space Treaty of 1976 \cite{ottr76}
declares that the Moon and other bodies ``shall be free for exploration'' and that the Moon is ``not subject to national appropriation by claim of sovereignty, by means of use or occupations''. Therefore, a Moon-based collider could potentially 
be built in the framework of a global effort, 
involving all, or most, of the countries on Earth. 
Such a project might also serve as a natural stepping stone towards future, larger space-based  infrastructures.

In this document, we examine the prospect and challenges of building a hadron collider around the $\sim$11000-km circumference of a great circle of the Moon, a Circular Collider on the Moon (CCM). A CCM could be made 100 times larger than the proposed FCC in the Lake Geneva Basin on Earth and would be six times larger than a CitS. Optimistically assuming a dipole magnetic field of 20 T \cite{botturasnowmass}, a CCM could potentially reach a proton-proton center-of-mass collision energy of 14 PeV, a thousand times more energetic than the LHC.

\newpage
\section{Construction and siting}\label{sec:site}

Two primary considerations for a CCM are its method of construction and its location on the Moon. The advantages, disadvantages, and costs of constructing a tunnel a hundred to several hundred meters under the surface, fully circumnavigating the Moon (``tunnelling", i.e., using lunar 
tunnel boring machines), vs. excavating a shallower trench and then covering it (``cut-and-cover") or constructing on the surface, would need to be studied, and an appropriate placement to be identified. Presented here are initial investigations as to the pros and cons of methods of construction and locations for a CCM. For future refinement, using, for example, the Unified Geological Map of the Moon~\cite{UGMM} along with further information expected from NASA's Artemis program~\cite{artemis}, as an input, the placement on the Moon could be optimised by applying the Tunnel Optimisation Tool, which was developed in the frame for the FCC study~\cite{Cook:2141838}.

\subsection{Method of construction}\label{subsec:construction}

Temperature extremes on the Moon present a challenge for a CCM. The dipole magnets needed to accelerate protons must be superconducting, which requires low temperatures compared to, e.g., room temperature on the surface of the Earth. When exposed to the sun, the temperature on the lunar surface can reach $\sim$127~\textdegree{}C~\cite{nasamoontemp} ($\sim$400 K), too high to consider any superconducting magnets. However, during Moon night, the temperature drops to about $-173$~\textdegree{}C ($\sim$100 K). This is approximately the temperature at which high-temperature superconductors (HTSs) could operate. This suggests one possible CCM scenario, Scenario A: Build the collider on the surface, charge an energy storage system during the lunar day, and operate the collider at night when the magnets are superconducting. The challenge with this scenario is that lunar days and nights last 13.5 Earth days each, leading to long stretches of non-operation. Scenario A has the benefit of avoiding the need to dig a tunnel, typically a large cost for any collider project.

Scenario A has a significant drawback because it would require a diameter $D$ smaller than twice the Moon radius $2R$ to ensure that the entire tunnel experiences Moon night at the same time. The fraction of the Moon night covering the entire collider can be estimated as $\left[ \pi- 2 {\rm arcsin} (D/(2 R)) \right]/\pi $, which vanishes for $D=2R$. The need to operate the CCM during lunar night immediately decreases the circular size, and the desire to operate for longer than, e.g., a few hours or one Earth-day, further reduces the size. Such reductions lead to a collider circumference significantly smaller than 11000 km and, hence, would lead to a smaller center-of-mass collision energy, reducing the attractiveness of the Moon as a site.

In addition to the above challenges, Scenario A also leaves the CCM exposed to not just extreme temperatures, but also to meteoroid strikes, which could potentially cause substantial damage to surface equipment. The Moon is bombarded by smaller meteroids regularly, resulting in numerous strong impacts each year that leave craters from a few centimeters to $\sim$10~m in diameter~\cite{spey2016}. (Impacts larger than this are rare.) A CCM constructed under Scenario A, i.e., on the lunar surface, would need to be adequately shielded to withstand impacts from small meteroids moving at up to 500~m per second~\cite{nasachurnurl}. Because other long-term research projects have been discussed to be located on the lunar surface~\cite{LCRTNASA}, such shielding should in principle be possible to achieve, or the damage from such strikes mitigated, though further study would be needed.

\begin{figure}[h]
    \centering
    \includegraphics[width=0.9\textwidth]{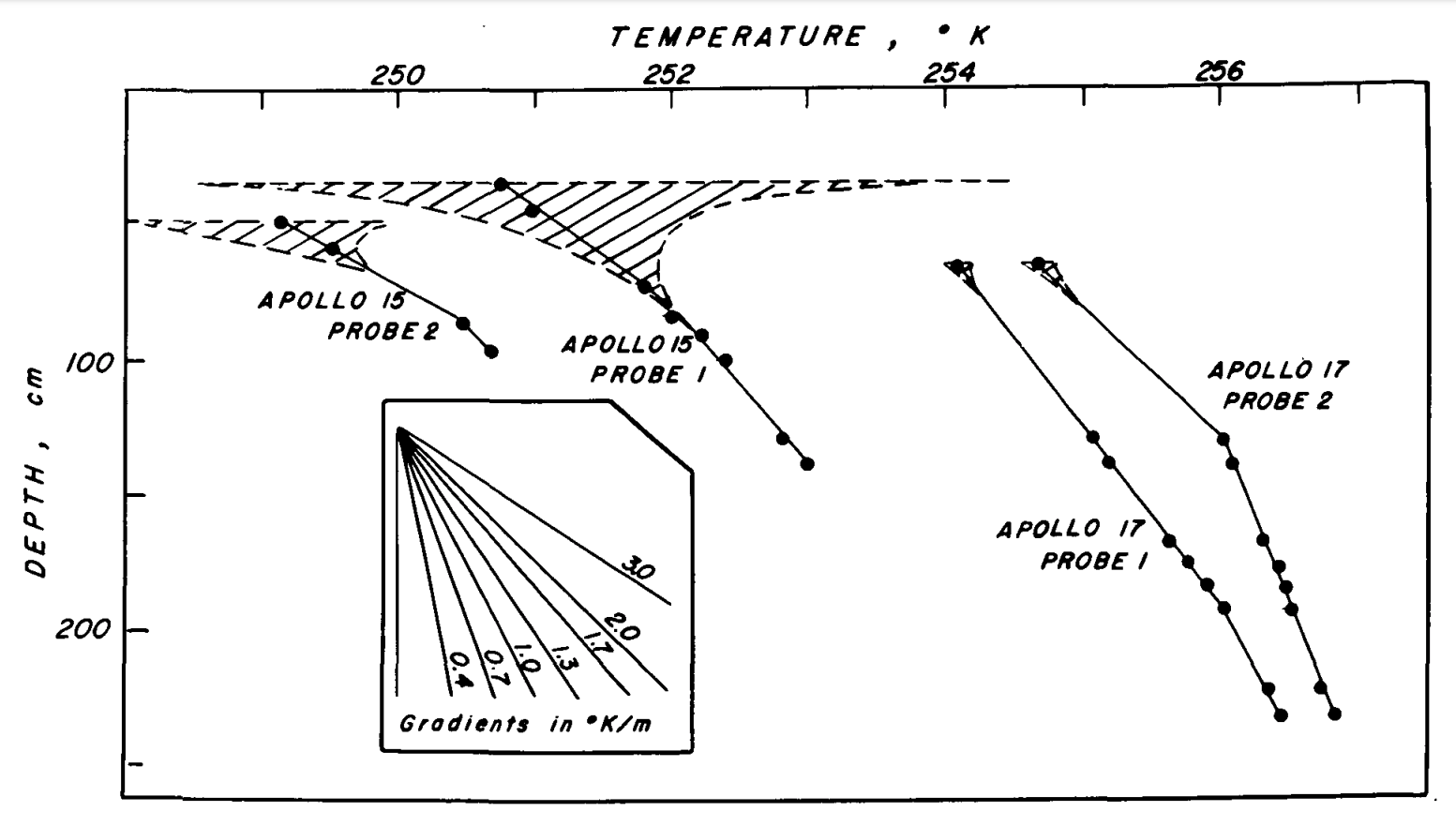}
    \caption{Temperature, in K, of the lunar subsurface, as measured by the Apollo 15 and 17 missions, taken directly from Ref.~\cite{1977NASSP.370..283L}. The striped regions for depths less than $\sim$70~cm show the variations in temperature measured throughout a lunar month. At depths exceeding $\sim$50~cm, temperature fluctuations due to lunar day-night cycles are negligible.}
    \label{fig:tempsubsurface}
\end{figure}

The drawbacks related to temperature variations and potential meteroid damage could be obviated by other scenarios. For example, in Scenario B a CCM could be constructed by digging and covering a trench a few meters deep (cut-and-cover). According to measurements from the Apollo 15 and 17 missions~\cite{1977NASSP.370..283L}, temperature fluctuations due to lunar day-night temperature cycles are negligible once one reaches $\sim$50~cm below the lunar surface, as shown in Figure~\ref{fig:tempsubsurface}.

However, the temperature of the lunar subsurface, while constant throughout lunar day and night, is still too high for presently known HTS materials, at 255--256~K at a depth of 1--2~m. Thus, a cooling system will be needed to maintain the HTS at a temperature of order 100~K and to enable CCM operation during lunar day and night for any scenario other than Scenario A. Additionally, it is possible that Scenario B by itself would not provide adequate shielding to protect against damage from meteoroid strikes, and that further surface shielding would need to be constructed anyway.

In a third scenario, Scenario C, a CCM tunnel could be bored a few tens to a few hundred meters under the lunar surface. Such tunnel boring would add substantial cost and time to the project and would still require a magnet cooling system, but would definitely prevent damage from (most) meteoroid strikes. Additionally, there has been much prior discussion of constructing tunnels for lunar habitats~\cite{mendell1985}, including the planned development of lunar tunnel boring machines~\cite{1988tamu.rept.....A,rostami2018,tunneltalk} operated remotely from Earth, and of the utilization of existing lunar lava tubes~\cite{2017Icar..282...47B,melosh2018}, and, thus, Scenario C would present key opportunities for collaboration with other organizations and initiatives.

\subsection{Placement and site}\label{subsec:placement}

The Moon has been measured to be a slightly scalene 
spheroid~\cite{garrick2014}, but for our purposes we model it as a sphere with a radius of 1738~km~\cite{lunarconst} and thus circumference of 10921~km, the equatorial circumference. More significant for a CCM is the fact that the lunar surface is rather rough. Local topographical deviations from an overall spherical shape can be large. The side of the Moon facing the Earth (the near side) is much smoother than the far side, where the largest deviations occur, with maximum elevations of $\sim$10.8~km~\cite{moonhigh} and minimum elevations of approximately $-9.1$~km~\cite{moonlow} relative to a reference sphere. (For reference, the highest mountain on Earth is Mt.\ Everest at an elevation of 8.8~km and the deepest part of the ocean is the Challenger Deep section of the Mariana Trench, at a depth of $\sim$11~km. However, Earth's radius is 3.7 times greater than the Moon's.) 

\begin{figure}[h]
    \centering
    \includegraphics[width=0.9\textwidth]{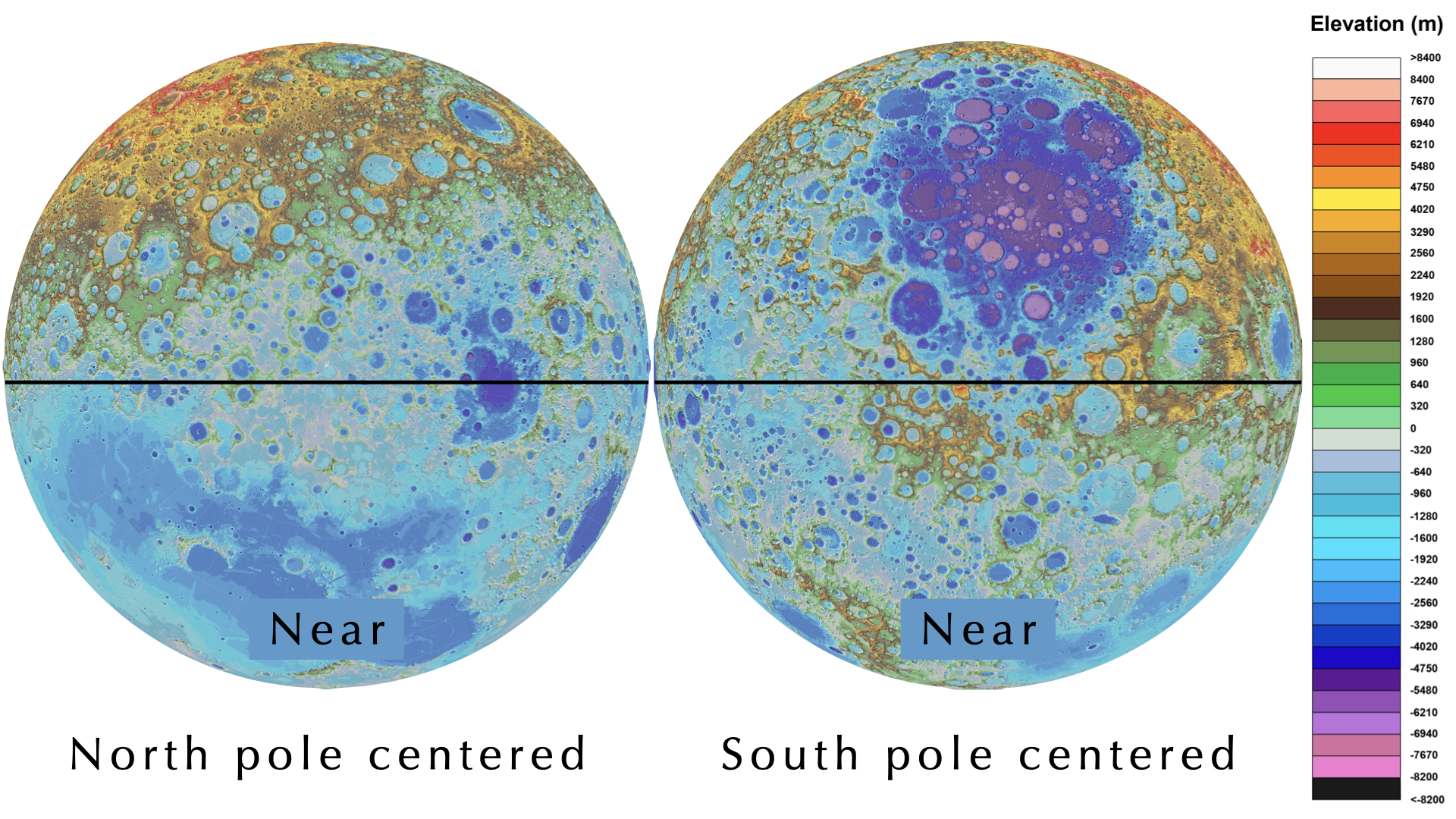}
    \caption{Schematic possible trajectory (black line) of a Circular Collider on the Moon (CCM) that could potentially avoid several major elevation changes, though not all. In the left image the north pole of the Moon is centered, while in the right image the south pole is centered. In each image the side of the Moon that faces the Earth is labeled ``Near''. Original topographical images were constructed with data collected by the Lunar Reconnaissance Orbiter Camera~\cite{2010SSRv..150...81R,2016SSRv..200..431H,2016SSRv..200..393M,2012ISPAr39B4..511S,LROC:EDR,2011JE003926,2010GL043751}.
    }
    \label{fig:CCMsite1}
\end{figure}

To ensure accessibility for maintenance and repairs, it will be beneficial to identify a CCM trajectory that avoids large elevation changes as much as possible. One such trajectory is shown schematically in Figure~\ref{fig:CCMsite1}. 

This trajectory circumnavigates the Moon approximately around the outer edge seen from the Earth, but slightly displaced toward the Earth. Such a trajectory appears to potentially avoid several major elevation changes, although not all. Irrespective of whether a tunnel a few hundred meters underground is bored (Scenario C) or a cut-and-cover approach is taken (Scenario B), the construction of a CCM will likely require some tunneling. Other major features, particularly the large {\it mare} (Mare Humboldtianum)~\cite{marehumboldt} in the northern lunar hemisphere, will likely require the CCM beam pipe and tubular shielding structure to be suspended, potentially several kilometers above the sunken surface. A detailed survey of this and other potential CCM sites is left for future work.

\clearpage

\newpage

\section{CCM machine parameters}
\label{sec:machine}

Since the strong and electromagnetic cross sections decrease as $1/\gamma^2$ (where $\gamma$ is the Lorentz factor of the accelerated protons),
the luminosity should increase with the square of the energy. 
Extrapolating from the FCC-hh (with a peak luminosity of about 
$3\times 10^{35}$~cm$^{-2}$s$^{-1}$ at $\sqrt{s} =$ 100 TeV), 
a 14 PeV collider should aim at the much higher luminosity of 
$6\times 10^{39}$~cm$^{-2}$s$^{-1}$.
The collider luminosity can be expressed as
\begin{equation}
L = \frac{f_{\rm rev} n_{b} N_{b}}{4 \pi \sigma_{x}^{\ast} \sigma_{y}^{\ast}} F_{\rm geom}
\label{lumi}
\end{equation}
with $\sigma_x^\ast = \sigma_y^\ast=\sqrt{\beta_{x,y}^{\ast}\varepsilon_{x,y}}$
the root-mean-square (rms) beam size at the interaction  point (IP), $n_b$ the number of bunches per beam, $N_b$ the bunch population, $f_{\rm rev}$ the revolution frequency, 
and $F_{\rm geom}\approx 1$ a geometric factor taking into account the so-called hourglass effect (variation of the beta function across the luminous region), crossing angle, etc. 
We can rewrite the luminosity by introducing the beam current
$I_{\rm beam} = e n_b N_b f_{\rm rev} $
and the beam-beam tune shift $\xi$
for round beams colliding head-on,
\begin{equation}
\xi = \frac{r_p N_{b}}{4 \pi \varepsilon \gamma}\; , 
\end{equation}
where $\gamma$ denotes the relativistic
Lorentz factor (here $\gamma\approx 7.5\times 10^6$
for 7 PeV protons, i.e., a 14 PeV proton collider), 
and $r_{p}$ the ``classical proton 
radius'' ($r_{p}\approx 1.5\times 10^{-18}$~m).
Equation (\ref{lumi}) becomes
\begin{equation}
L = \gamma\;  \left(\frac{1}{e r_{p}}\right) \;  
\frac{I_{\rm beam}}{\beta^{\ast}}
\xi F_{\rm geom}\; , 
\label{lumi2}
\end{equation}
with $\beta^{\ast}\equiv \beta_{x}^{\ast}=\beta_{y}^{\ast}$.

From (\ref{lumi2}),  
a linear increase with energy will be
achieved naturally, thanks to the adiabatic
damping of the proton beam's transverse 
geometric rms emittance $\varepsilon$, 
for a constant normalized emittance ($\gamma \varepsilon$). 

Fortunately, 
since the transverse radiation damping time is short, of order 30 s (compared with 1 day at the LHC), the emittance will shrink quickly until it is 
balanced by an external 
transverse noise excitation that will 
need to be applied in order  
to maintain stable collisions at a still acceptable value of the 
beam-beam tune shift \cite{PhysRevSTAB.18.101002}.

For the CCM, 
the transverse  damping decrement $\lambda_d =  U_0/(2 E_{\rm beam}n_{\rm IP})$ is about  $1.2\times 10^{-3}/n_{\rm IP}$ (200 turns damping time),
where $n_{\rm IP}$ designates the number of interaction points.
The value of the damping decrement is reminiscent of the e$^+$e$^-$ collider LEP, where the following approximate scaling law was established
\cite{Assmann:453821}: 
\begin{equation}
    \xi ^{\rm max} \propto \lambda_{d}^{0.4} \; . 
\end{equation}
From Ref.~\cite{Assmann:453821}, we can expect a maximum  
tune shift $\xi$ between 0.05 and 0.08, depending on the number 
of interaction points. Recent colliders, e.g., KEKB, have achieved
higher tune shifts still.  
Conservatively assuming a maximum tune shift of 0.06 
yields another factor of 10 in luminosity, compared with FCC-hh.  
We then arrive at an unprecedented luminosity of 
$2\times 10^{38}$~cm$^{-2}$s$^{-1}$,
which nevertheless falls short of our ideal 
target by a factor $\sim$30. 
Higher tune shift or smaller IP beta functions would 
further boost the luminosity.

Based on the above considerations,
Table~\ref{param} presents tentative proton-collider parameters for the CCM,
and compares them with those of the planned FCC-hh and the HL-LHC.
The CCM beam energy follows from $E_{\rm beam}=ceB_{\rm dip}F_{\rm dip} C/(2 \pi)$,
where $c$ is the speed of light, $e$ the electron (proton) charge,
$C\approx 11,000$~km  the ring
circumference, $B_{\rm dip}=20$~T the dipole field, 
and $F_{\rm dip}\approx 0.67$ the dipole filling factor, 
similar to the LHC's. 
Beam current and bunch spacing are kept roughly 
the same as for the FCC-hh and the present LHC.
Assuming that the CCM will operate at $\xi=0.06$,
the rms beam size at the interaction point (IP) 
is 120 nm, more than two times larger than the vertical
IP beam size at SuperKEKB.

\begin{center}
\begin{table*}[ht]
\hfill{}
\begin{tabular}{|l|c|c|c|}
\hline
\hline
Parameter &
\multicolumn{1}{c|}{CCM} & 
\multicolumn{1}{c|}{FCC-hh} & 
\multicolumn{1}{c|}{HL-LHC}   \\
\hline
Maximum beam energy $E_{\rm beam}$ [TeV] &
7,000  & 50 & 7 \\
 Circumference $C$ [km] & 
 \multicolumn{1}{c|}{11,000}
 &
  \multicolumn{1}{c|}{97.8} &
   \multicolumn{1}{c|}{26.7} \\
 Arc dipole magnet field $B_{\rm dip}$ [T] & 
 \multicolumn{1}{c|}{20}
 &
  \multicolumn{1}{c|}{16} &
   \multicolumn{1}{c|}{8.3} 
 \\ 
\hline
Beta function at interaction point (IP) $\beta_{x,y}^{\ast}$ [m] &
0.5 & 0.3  &  0.15 
\\
Transverse normalized~rms emittance $\varepsilon_{n}$ [$\mu$m] & 0.2 & 2.2  &  2.5    
\\
Rms interaction-point beam size [$\mu$m] & 0.12 & 3.5  & 7 
\\
Beam current [A] & 0.5 & 0.5 & 1.12 
 \\ 
Bunches per beam $n_{b}$ & 1,200,000 & 10,400 & 2,760  \\
Bunch spacing [ns] & 25 & 
25 & 25  \\
Bunch population $N_{b}$ [$10^{11}$] & 1.0  &  1.0 & 2.2  \\
Energy loss per turn $U_{0}$ [MeV] & $1.7\times 10^7$ & 4.67 &  0.007
\\ 
Synchrotron radiation power $P_{\rm SR}$ 
[MW] & $8.5\times 10^{6}$ & 4.8 & 0.014 \\
Critical photon energy $E_{\rm cr}$ [keV] & 105,000 & 4.3 &
0.044 \\
Transverse~emittance~damping time $\tau_{x,y}$
[h] & 0.004 & 1.0 & 25.8 \\
Beam-beam parameter per interaction point, $\xi$ [$10^{-3}$] 
& 60 & 5.4 & 8.6 
\\ \hline
Luminosity per interaction point $L$ [$10^{34}$~cm$^{-2}$s$^{-1}$] &
$\sim$20,000 & $\sim$30 & 5 (leveled)
\\
  \hline
Number of events per bunch crossing (pile-up) &
$\sim$10$^{6}$ & $\sim$1000 & 135 
\\ \hline
Maximum integrated luminosity per experiment [ab$^{-1}$/y] & 
$\sim$2000 & 1.0  & 0.35   \\ 
\hline\hline
\end{tabular}
\hfill{}
\caption{Tentative proton-proton  parameters for
CCM, compared with FCC-hh 
and HL-LHC \protect\cite{bordry2018machine}.}
\label{param}
\end{table*}
\end{center}

Quantities of interest are the beam lifetime, the synchrotron radiation,
and the power consumption.
The beam lifetime is \cite{PhysRevSTAB.18.101002} 
\begin{equation}
\tau _{\rm beam} = \frac{n_{b}N_{b}}{L n_{IP}\sigma_{\rm tot}}\; ,  
\label{lifetime}
\end{equation}
where the total cross section $\sigma_{\rm tot}$ can be approximated
as \cite{Menon:2012dr}
\begin{equation}
    \sigma_{\rm tot}\; [{\rm mbarn}] \approx 42.1 s^{-0.467}
    -32.19 s^{-0.540} + 35.83 + 0.315 {\rm \ln}^{2} \left( \frac{s}{34} \right) \; ,
\end{equation}
and $s$ designates the square of the center-of-mass energy
in units of GeV$^2$.  
For the collision energy of the CCM we find 
 $\sigma_{\rm tot} \approx 283$~mbarn. 
With $n_{\rm IP}=2$ IPs, the beam lifetime (\ref{lifetime})
is only about 18 minutes.

Therefore, we could require rapid top-up injection in order to maintain a constant beam current.
For example, injecting every 100 seconds would keep the beam current constant to within 10\%.

The integrated luminosity can now be estimated under assumptions similar to those adopted for 
proposed future colliders at CERN \cite{bordry2018machine}.
With 160 days scheduled for physics per year,
and assuming a 70\% availability of the complex,
we estimate an integrated luminosity of $\sim$2000~ab$^{-1}$ per experiment and per year. 

Critical photon energies in the arcs, $(3/2) \hbar c \gamma^3 / \rho $ with
$\rho \approx F_{\rm dip}C/(2 \pi)$,  
reach the level of 100~MeV. Fortunately, in the excellent vacuum on the Moon no beam pipe would be needed. With a residual molecular or atomic density of order $10^{12}$/m$^{3}$ or less~\cite{MFS}, the vacuum quality on the Moon surface is at least 100 times better than the vacuum in the beam pipes of the LHC of $10^{15}$~H$_{2}$/m$^{3}$~\cite[Chapter 12]{Bruning:782076}. Therefore, the charged particles of the CCM could circulate without a dedicated vacuum system, shielding, beam pipe, and pumping system, resulting in a great cost saving. Open plane dipole magnets would allow the photons to escape into the tunnel wall or into the Moon sky, depending on the scenario.

\section{Powering}\label{sec:powering}

A prominent concern for the CCM is identifying a power source substantial enough to both sustain the machine and compensate for energy lost due to synchrotron radiation. 
The energy loss per turn increases as $\gamma^4$ and decreases with circumference $C$ (or radius) only inversely linearly, as $1/C$.
It also rises linearly with beam current.
Doing the algebra, and starting from 4.8 MW emitted at the FCC-hh,
we arrive at the interesting  
number of 8,500 GW or 8.5 TW.
This power must be sustained by a distributed radiofrequency system, or by an alternative acceleration mechanism.
A power source at the level of 10 TW would thus appear to 
be required. For comparison, the total energy consumption for the entire Earth in 2019 corresponded to a source of $\sim$18--20 TW~\cite{globalpower}.

Fortunately, a number of possible powering scenarios for lunar habitats and other projects on the Moon have already been explored, since a major consideration of any project proposed for the Moon is power. (See, for example, Ref.~\cite{CLIMENT2014352}.) Some combination of these could in principle be utilized for a CCM.

For example, NASA and the U.S. Department of Energy have begun studies of fission surface power~\cite{nasafission,doefission1,doefission2} to be deployed for small-scale projects with minimal power needs on the Moon~\cite{mitnuclear}. Since such studies target power outputs in the kW range, these are, however, 
unlikely to be useful for a CCM as currently envisioned.

An alternative option would be the construction of full-scale fission reactors on the Moon like those on Earth. Even if this proved possible, it is unclear whether the power output could meet the needs of a CCM. As of this writing, there are 444 operable fission power reactors used for non-military purposes in the world, with a combined power output of 394 GW~\cite{iaea}. Achieving tens of TW with existing fission technology will, thus, be potentially challenging.

Fusion power is in principle more promising, but there are challenges associated with fusion reactors, as well. The target power output of ITER~\cite{iter} is 500 MW, from 50 MW of injected power, and, thus, a CCM would require a power output equivalent to $\sim$20k ITER reactors. Clearly fusion power technology will advance significantly in the next several decades, but it remains unclear when it could be harnessed to reach the 10s of TW level. As a result, alternative powering methods for the CCM should be examined.

More promising for a CCM is power from the sun. Due to its lack of atmosphere, solar power would be plentiful on the Moon. Solar power on the Moon has the additional benefit that the materials needed for manufacturing photovoltaic cells are likely already present in the rocks and dust of the Moon (as discussed in, e.g., Ref.~\cite{1367615}). One possible powering solution for the CCM could, thus, be a large-scale harnessing of solar power in the form of the construction of a small patch of a Dyson sphere~\cite{Dyson1667} around the sun, on or near the Moon. The total power emitted by the sun is about 400 yottawatt ($4 \times 10^{26}$ W)~\cite{NSFS}. Therefore, less than a thirtieth of a part per trillion of the sun power would be needed to operate the CCM. Indeed, considering the distance between the Moon and the sun of about 150 million km, the power required corresponds to $\sim$0.07\% of the sun power incident on the Moon surface. Assuming 100\% efficient solar panels, a CCM would thus require the construction of a solar power array of $\sim$6.7$ \times 10^{3}$ km$^{2}$ on the lunar surface, an area slightly larger than the size of the U.S. state of Delaware.
Current solar power technology achieves $\sim$20\% efficiency on Earth, and so the total array on the Moon would need to be larger. Solar power on the Moon will also likely be complicated by the fact that radiation exposure and the high temperatures during lunar day can severely degrade and damage solar panels. NASA is currently studying ways to address these challenges~\cite{nasasolar}.

\begin{figure}[h]
    \centering
    \includegraphics[width=0.5\textwidth]{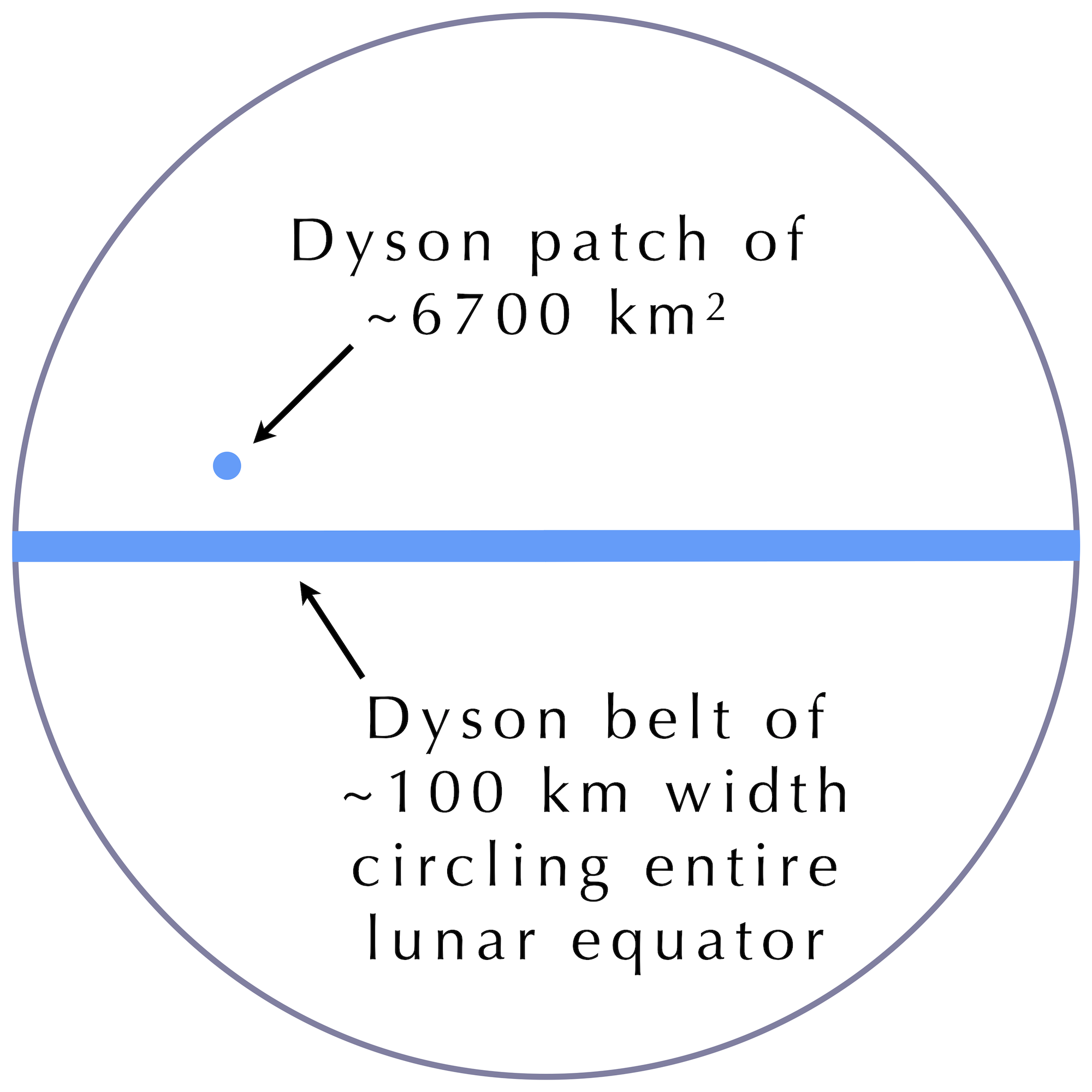}
    \caption{Schematic demonstration of the relative size on the Moon (large gray circle) of a Dyson patch of $\sim$6.7$ \times 10^{3}$ km$^{2}$ (small blue dot) that could potentially collect enough solar power to provide the $\sim$10~TW estimated to be needed by a CCM, assuming the patch could be constantly in sunlight. The patch is slightly larger than the size of the U.S. state of Delaware. Because such a spot that is constantly exposed to the sun does not exist on the Moon, also shown is the size of a Dyson belt of 100 km in width, circumnavigating the Moon around the lunar equator. Such a Dyson belt would always have a portion of the solar panels exposed to the sun. The concept is similar to the Luna Ring concept proposed by the Shimizu Corporation~\cite{shimizu}.}
    \label{fig:CCMDyson}
\end{figure}

The other major limitation related to using solar power in the form of a Dyson patch is that there is no part of the Moon that is constantly exposed to the sun. A solution to address both the issue with solar panel efficiency for a CCM and the issue with sun exposure would be to build a Dyson band or belt around the equator of the Moon to continuously collect sun power. Both a Dyson patch of $\sim$6.7$ \times 10^{3}$ km$^{2}$ and a Dyson belt of 100 km in width are shown schematically in Figure~\ref{fig:CCMDyson}.

Such a Dyson belt would additionally easily provide ample power to multiple lunar projects, as well as potentially send power back to Earth. Such an idea was championed by David Criswell for many decades~\cite{criswellLSP} and has been recently studied in detail by the Shimizu Corporation with the Luna Ring concept~\cite{shimizu} and by the Japan Aerospace Exploration Agency (JAXA) 
for the Space Solar Power Systems (SSPS)~\cite{SSPS}.

\newpage
\section{Injector}\label{sec:inj}
\begin{figure}[htbp]
    \centering
    \includegraphics[width=0.7\textwidth]{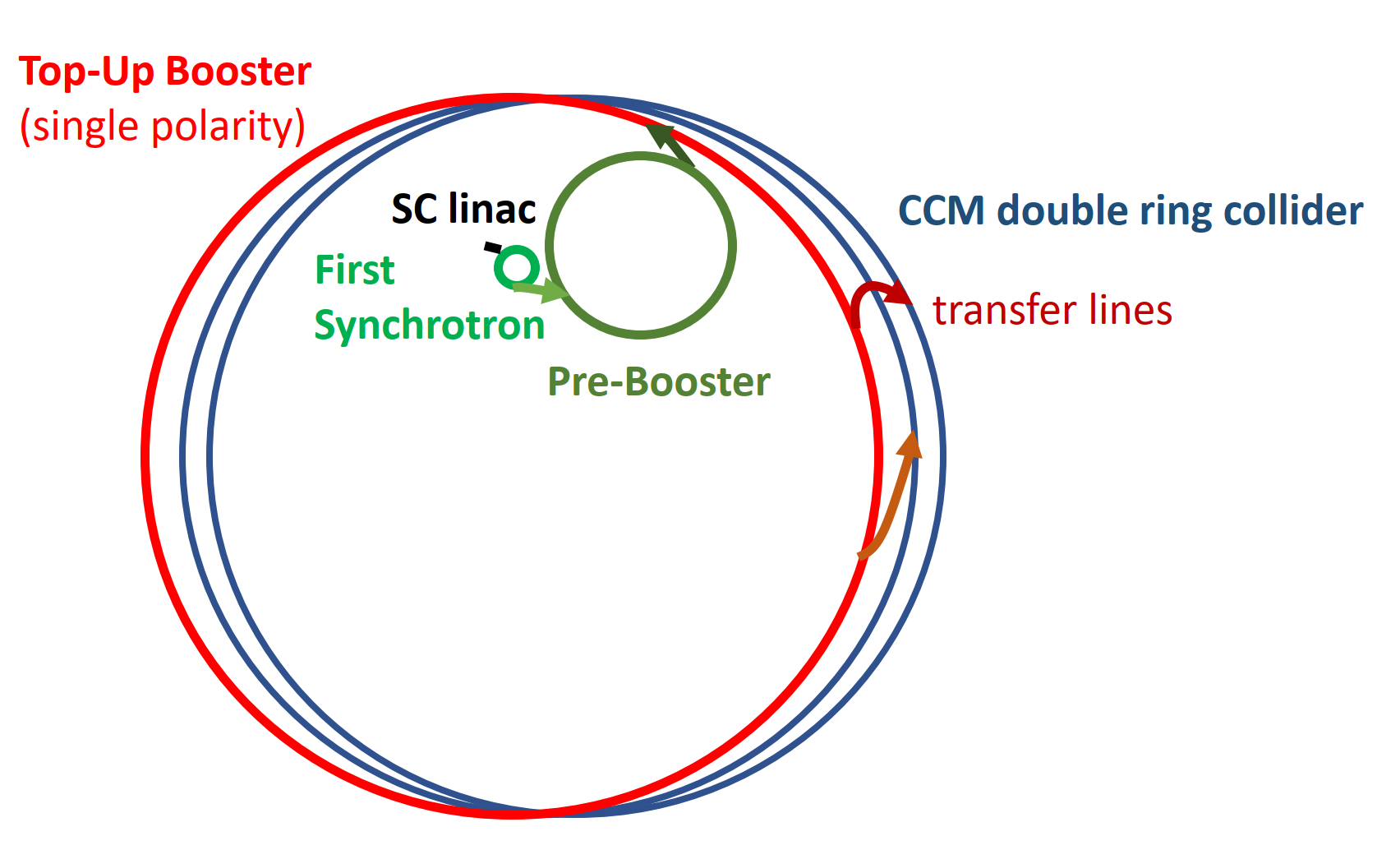}
    \caption{Sketch of a possible CCM injector chain, with 
    parameters as listed in Table \protect\ref{tabinj}. The ``First Synchrotron'' circle (light green) is twenty times larger in circumference than the LHC.
    }
    \label{fig:inj}
\end{figure}

For the design beam lifetime of 18 minutes,
a top-up injection at time intervals of 100 s for each ring 
would assure that the beam current stays approximately 
constant, within $\pm 5\%$ of the average value,
translating into fairly constant luminosity and acceptable
charge-imbalances between colliding bunches. 
This top-up injection at a rate of 0.02 Hz for the two beams   could be achieved with a rapid cycling 
``conventional'' full energy booster
(though achieving the implied 
ramp rate with 20 T magnets might be non-trivial).
The full-energy booster injector must have the same size and the same peak dipole field of 20 T as the collider.  
The booster can be preceded by a series of 
lower-energy and perhaps lower field 
circular accelerators. Each synchrotron could increase
the beam energy by about a factor of 20,
as for example is done with the PS, SPS, and LHC at CERN.
Field is balanced with the circumference and number of cycles
required to fill the subsequent machine.
As cycle times for the lower-energy machines become rather 
short, injecting into the first synchrotron of the chain
at an energy of around 1 TeV, from a superconducting proton linac,  could be an interesting option.
Table~\ref{tabinj} presents parameters of 
an example configuration sketched in Fig.~\ref{fig:inj}.
It is evident that the complex could be constructed in stages,
and both the proton linac and the first synchrotron be initially 
used for other applications, while the downstream accelerators are
being constructed. 

Incidentally, the high-power linac could also be an excellent 
proton driver to produce the muons for a muon collider. (See Ref.~\cite{muoncollider2020} and references therein for a recent discussion of potential muon colliders.) 
If constructed on the Moon,  
the collision energy would no longer be limited,   
by the neutrino radiation hazard,  
to a few tens of TeV~\cite{King:1999gb}, 
as an earth-based muon machine would be.

\begin{table}[htbp]
\caption{Parameters for a possible CCM injector chain.}
\label{tabinj}
\centering 
\begin{tabular}{lcccc}
\hline\hline
Synchrotron & Circumference [km] & Max.~Dipole field 
[T] & Cycle time [s] & Extr. / Inj. Energy\\
\hline
Top-Up Booster & 11,000 & 20 & 50 &  7 PeV / 350 TeV \\
Pre-Booster & 2,750 & 4 & 12.5 & 350 TeV / 17.5 TeV \\ 
First Synchrotron  &  550 & 1 & 2.5 & 17.5 TeV / 0.9 TeV \\
Superconducting linac & 50 (length) & --- & CW & 0.9 TeV / $\sim$0 \\ 
\hline\hline
\end{tabular}
\end{table}

\newpage
\section{Dipole magnets}\label{sec:dipoles}

An important consideration for the CCM is the identification and sourcing of sufficient raw materials to constitute the dipole magnets needed. In contrast to the more than 1200 dipole magnets of the LHC and the planned $\sim$5000 dipole magnets of the FCC, each about $14$ meters long, the CCM will require $\sim$5.5$ \times 10^{5}$ dipole magnets for the full, 11000-km accelerator alone, assuming a similar magnet length. (Additional magnets will be needed for the supporting accelerators in the injector complex; for simplicity we consider the CCM itself in the following.)

As discussed in Section~\ref{subsec:construction}, the dipole magnets of the CCM will likely be high-temperature superconductors (HTSs) targeting a magnetic field of at least $\sim$20~T. Two HTS approaches and technologies present themselves as possibilities for the CCM.

The first is to use HTSs based on tapes of rare-earth elements (REEs) combined with barium-copper-oxide, collectively referred to as ReBCO. (See Ref.~\cite{vanNugteren:2018xal} for a recent review of ReBCO technology and prospects for use in magnets for future colliders.) In practice, HTSs for accelerators are composed of both superconducting material and some amount of stabilizer, such as copper. In the case of ReBCO-based HTSs, the ReBCO layer is a small fraction of the entire material, $\sim$2--3\%, and the REE component is $\sim$30--40\% of the ReBCO, a comparably small fraction of the total. A sufficient quantity of usable REEs such as yttrium or gadolinium would need to be identified for the dipoles of the CCM, presenting a challenge because, as their name implies, such elements are rare. Moreover, a 1--2~$\mu$m silver layer 
might be needed on top of, or underneath, 
the ReBCO HTS material in addition to 50--100~$\mu$m copper
stabilizer~\cite{Bonura_2014}.

As a benchmark, we consider a CCM constructed of a series of 14-m dipoles, each of which consists of 2 tons of conductor total. This corresponds to $\sim$24 kg of REE needed per dipole, and a corresponding total amount of REE needed for the CCM of $\sim$1.32$ \times 10^{7}$ kg, or $\sim$13200 tons.

However, hybrid magnets could be constructed from some combination of ReBCO-based HTS as an inner high-field layer and other, more conventional materials, such as Nb$_{3}$Sn, envisioned for the FCC-hh, or Nb-Ti, as for the LHC, for the outer lower-field layers.
For example, a 20 T hybrid magnet consisting of three layers was considered in Ref.~\cite{Assmann:1284326}.   
Under the assumption of CCM dipole magnets consisting of, e.g., 50\% Nb$_{3}$Sn and 50\% ReBCO, and given the benchmark above, this corresponds to $\sim$6600 tons of REE needed.

The total amount of all REEs on Earth is estimated to correspond to 120 million tons~\cite{globalREE}. Sourcing REEs for the CCM on Earth and then transporting them to the Moon could, in principle, be feasible, though Earth-to-Moon material
transport should likely be avoided or kept to a minimum due to the cost involved. Sourcing REEs on the Moon, instead, would alleviate the need for transport. However, currently it is unclear whether REEs are plentiful enough on the Moon to be collected and used. Recent estimates suggest that lunar REEs could be abundant, but it remains to be confirmed whether they would exist in usable amounts for a CCM. (See, e.g., Ref.~\cite{resources6030040}.) In principle, however, irrespective of their source, $\sim$6600 tons of REE could be collected over the decades prior to and during the construction of the CCM.

Moreover, ancillary material such as copper or silver will need to be sourced, as mentioned above. Evidence from lunar missions indicates the possible presence of silver on the Moon, as described in, e.g., Ref.~\cite{Schultz468}, along with oxygen, hydrogen, sodium, and other elements, but the question as to whether such elements are present in quantities sufficient for ReBCO-based HTSs for the CCM remains open.

The second option would be to deploy magnets that utilize iron-based superconductor (IBS)~\cite{ironhts}, as is foreseen to be used for the 12--24~T magnets of the SPPC. IBSs based on iron and 
pnictides (typically arsenic (As) and phosphorus (P))  
would alleviate the need to source thousands of tons of REE, instead requiring the sourcing of something on the order of a million tons of iron, arsenic and phosphorus. 
The total amount of usable metals and other elements (including iron
and pnictides) on the Moon is the subject of much study, 
with current estimates suggesting that accessible iron could be plentiful~\cite{crawford2014lunar,HEGGY2020116274}. Further studies of the lunar composition will be needed to determine whether sufficient iron and pnictides 
for a CCM could be sourced on the Moon.

Magnet R\&D for IBS technology is developing~\cite{Zhang_2021,Wang_2019,doi:10.1142/S0217751X19400037,Tang2017} and IBS could likely prove an appealing alternative to ReBCO-based HTS.
A rough price estimate for the SPPC IBS magnets is 200k RMB (or $\sim$28k CHF as of May 2021) per meter, roughly shared between the cost of the IBS conductor and the mechanical components of the magnet~\cite{QingjinXu}.
Assuming the same cost, the dipole magnets of the CCM would amount to about 200 billion CHF, not including any possibly required material transport from Earth to Moon. Such cost estimates will likely be greatly affected by cooperative mining and manufacturing systems on the Moon that could serve multiple projects, including the CCM, potentially lowering the overall magnet expenses.

A more extensive study of the availability of the materials needed and construction costs for the magnets of the CCM and the supporting injector complex is left for future work.

\newpage
\section{Duration of construction, installation, and commissioning}\label{sec:time}

As discussed in Section~\ref{subsec:construction}, Scenario C --- which involves excavating tunnels a few dozen to a few hundred meters under the lunar surface to avoid temperature variations, radiation damage to workers and facilities, and meteoroid strikes --- likely presents the best option for the CCM. The completion of the tunneling complex needed for both the main, $\sim$11000-km tunnel of the CCM and the additional tunnels needed to house the other accelerators of the injection complex (as described in Section~\ref{sec:inj}) --- as well as the shafts from the lunar surface to the CCM tunnel, auxiliary tunnels, caverns, and alcoves --- could target a seven-to-ten-year schedule. Such a tunneling schedule relies upon the construction, availability, and operation of a number of lunar tunnel boring machines (LTBMs) working simultaneously and in parallel and, thus, relies upon technology mature enough to allow for remote handling.

Construction of manufacturing centers, factories, underground buildings for support personnel, data centres, communication systems, and various infrastructure pieces, as well as construction of the dipole magnets and CCM machine components, could also proceed in parallel with tunnel excavation. As tunnel excavation would necessarily displace a large amount of rock from the lunar regolith, such resources could potentially be used in the construction of the accelerator components or detectors, and could be made available to other projects on the Moon.

Additional time --- perhaps five years --- will be needed for installation of the machine, commissioning, and the construction and installation of the detectors, which would bring the total time from the beginning of construction to the beginning of data-taking to about 15 years.

Prior to the beginning of CCM construction, several broader developments must occur. Reliable systems for transporting humans and some resources between the Earth and the Moon would need to be robust, regular, and affordable. Remotely-operated LTBMs must be mature, industry-quality, and available, something that may be feasible by the 2040s, depending upon several factors~\cite{rostamipriv}. Detailed studies of the geological composition of the lunar regolith, as well as surveys of the lunar subsurface, to ensure the accessibility of sufficient resources to construct all components of the CCM, would need to be performed. A powering system such as the Dyson belt / Luna Ring solar powering system discussed in Section~\ref{sec:powering} would need to be constructed; under the assumption that such a project would begin construction by 2035 at the soonest (according to a projection by the Shimizu Luna Ring team), it could potentially be able to provide power to other lunar projects by the 2040s or 2050s. Underground facilities for longer-term stays by researchers would need to be built and installed.

Given the multiple technological unknowns and large uncertainties associated to them, it is challenging to arrive at a concrete estimate as to the timeline for a project such as the CCM. 
However, the uncertainties and unknowns associated to each of the needed technological advancements vary widely in size and scope. Some of the needed advancements are being currently pursued and are clearly foreseen for the future, while others are decidedly more ambitious and the timelines more speculative. For example, in a few decades from now, 20-T magnet technology can be taken for granted~\cite{botturasnowmass}, whereas it is much more challenging to assign a concrete timeline to the availability of, e.g., remotely-operated LTBMs. The time scale for a CCM is not defined by accelerator technology (because such development continues and will continue apace here on Earth, irrespective of any potential utilization on the Moon), but rather by the advances in the construction of lunar habitats, lunar mining and production capabilities, and suitable large-scale solar energy generation.

The NASA Artemis program~\cite{artemis} aims at establishing a Base Camp on the Moon surface by the end of the 2020s, and a ``Gateway'' in lunar orbit. These elements will support robots and astronauts in 
explorations and science missions. 
In parallel, China and Russia plan the joint construction of an  International Lunar Research Station (ILRS)~\cite{spectrumilrs}, which, similar to Artemis, will include both a space station in lunar orbit  
and a Moon base on the surface.  
The ILRS is expected to become operational a few years after the Artemis base, namely from 2036 onwards, 
and to provide a range of scientific facilities and equipment for studying lunar topography, geomorphology, chemistry, geology and internal structure of the Moon~\cite{ilrsspace}. The ILRS will explore and advance    
the potential extraction of helium-3 
from the Moon's surface,  and 
the 3D printing of solar panels at the Moon's equator. 
The lunar solar panels would capture solar energy, 
that could be transmitted to Earth by lasers or microwaves, or, more easily, 
be consumed directly on the Moon~\cite{spectrumilrs}. 

Various encouraging studies and forecasts 
for the second half of the 
21st century also exist. 
For example, 
working groups at the NASA's Marshall Space
Flight Center (MSFC) developed estimates of the operations 
and economics of space activities in the vicinity of the Earth 
and Moon in the 2050 time frame,
along with a picture for 2100~\cite{7119255}. 
A scenario for the 2050 ``cis-lunar econosphere'' 
identifies In Situ 
Resource Utilization (ISRU) and on-site, 3D 
manufacturing as critical technologies. Other MSFC predictions or assumptions 
are that, ``Mining operations on the moon’s surface ... will be 
robotic in nature, with human oversight", and that   
a 300-person habitat in the Moon’s vicinity, ``will 
house personnel to run mining operations, way stations, 
and construction of a large habitat'' on the Moon itself.  
A report from the 
US Air Force Space Command, published in 2019, 
predicts that, in a 
positive scenario, by 2060, 
``Multiple nations have [M]oon bases or colonies competing ... 
in providing key 
infrastructure for commercial exploitation of the [M]oon and continued development of manufacturing 
and facilities across cislunar space''~\cite{usaf}. 

Given these plans, and under the assumption that the already excellent progress on the necessary base technologies, e.g.~lunar mining, lunar tunneling, in-situ fabrication, and lunar/spatial energy production, continues as predicted, it is reasonable to estimate that the CCM could begin initial construction at the earliest in the 2070s or 2080s.
It is therefore to be expected that the CCM might start producing data not before the beginning of the 22nd century, if not much later.

\newpage
\section{Other considerations and opportunities}\label{sec:other}

There are several important aspects, considerations, and opportunities associated with and presented by the CCM that are left for future work. The following is a partial list of these aspects.

\begin{itemize}
    
    \item \textbf{Detectors:} As discussed in Section~\ref{sec:machine}, the beam lifetime of the CCM is already very short, at around 18 minutes. This suggests two interaction points (IPs) and, hence, two detectors. A larger number of IPs would require a more powerful injector, which is a challenging prospect, given the conditions discussed. Two IPs and two detectors suggests that the detectors both be designed in a general-purpose manner, similar to the way ATLAS and CMS were designed at the LHC. However, detector technology and hardware needed for operation under the extreme conditions of the CCM (e.g., performing tracking and triggering when the average number of events per proton bunch crossing is on the level of 10$^{6}$) would need to be developed. Construction of the detectors would need to be conducted on the Moon, to avoid costly and potentially damaging transport from the Earth to the Moon, and could proceed within the same facilities built for the construction of the CCM accelerator complex.
    
    \item \textbf{Data management and transfer:} The collision data produced at the CCM and collected by the detectors would likely be initially stored at a data centre located on the Moon itself. Initial reconstruction of the data could take place at the lunar data centre before being transmitted to multiple data centres on the Earth for redundant storage and further analysis by physicists, similar in spirit to the Worldwide LHC Computing Grid~\cite{wlcg}. In the event that a sizable percentage of the CCM physicists were based on the Moon (see below), an alternate, Moon-based grid (a Lunar CCM Computing Grid) system could be employed, e.g., if habitats were constructed within multiple lunar lava tubes. Data transfer between the Moon and Earth should be relatively straightforward to accomplish, at high rates, given the advanced state of current deep space communications research at, e.g., NASA~\cite{NASASCAN}, the ESA~\cite{ESAcomm}, and JAXA~\cite{JAXAcomm}, and the likely major progress in space communications in the next 50 years.
    
    \item \textbf{Physics program:} The physics program for the detectors at the CCM would likely have three pillars. The first would be a general search program for new particles and fields, including any deviations from Standard Model (SM) expectations as well as more specific, model-dependent searches for beyond the SM phenomena at an unprecedented center-of-mass proton-proton collision energy of 14 PeV. The second would be precision study of SM predictions and properties at such high energies. The third pillar of the CCM research program would be informed by what is observed by preceding collider projects such as the FCC, the SPPC, or the CitS.
    
    \item \textbf{Heavy ion collisions:} The physics program of the CCM could --- and of course should --- be extended to include heavy-ion collisions in addition to proton-proton collisions, at the equivalent centre-of-mass energy. 
    The searches for new physics in heavy-ion collisions carried out at the LHC~\cite{lhcions}, and envisioned for the FCC-hh~\cite{fcchhions}, could be expanded into a 
    much higher energy range.  
    
    \item \textbf{Muon collider:} As mentioned in Section~\ref{sec:inj}, the high-power linac of the CCM could also be an excellent proton driver to produce the muons for a muon collider. (See Ref.~\cite{muoncollider2020} and references therein for a recent discussion of potential muon colliders.) If constructed on the Moon, the collision energy would no longer be limited, by the neutrino radiation hazard, to a few tens of TeV~\cite{King:1999gb}, as an earth-based muon machine would be.

    \item \textbf{Detectors as cosmic ray observatory:} The detectors of the CCM could in principle be used as cosmic-ray observatories during commissioning and non-operation.
    
    \item \textbf{Remote operation vs. on-site intervention:} The prospect of thousands of CCM maintenance staff members and researchers residing permanently on the Moon is sub-optimal, owing to the likely costs associated to travel between the Earth and the Moon. Thus, a full examination of the needs of a largely remotely-operated accelerator and detector complex needs to be conducted. However, given the complexity of both the CCM accelerator complex and the two detectors --- and the fact that current high-energy physics projects require significant manual operation and occasional intervention --- it's probable that some number of physicists, technicians, engineers, and support staff for the CCM will be stationed on the Moon. In the interests of keeping this number to a minimum, as mentioned above, a full study of the operational needs of the CCM will be necessary.
    
    \item \textbf{Lunar Particle and Astrophysics Research Center (LPARC):} The construction of the CCM will require global collaboration similar in spirit to that of the LHC and FCC at CERN, but much larger in scale. Because of its remote location and unique circumstances, the CCM could be constructed as one component of an entire Lunar Particle and Astrophysics Research Center (LPARC). In addition to the CCM, an LPARC could feature multiple other projects and experiments such as observational astronomy~\cite{LCRTNASA}, gravitational wave detectors, a very-long-baseline neutrino detector between Earth and Moon~\cite{1992LPI....23.1537W}, dark matter experiments, precision measurements (in low-gravity conditions), quantum computing, a quantum technology center, a large data centre (see above), etc. Many such projects have been proposed over the years and all could benefit from collaboration and synergy with the CCM.
    
\end{itemize}

\newpage
\section{Classification of unknowns}\label{sec:unknowns}

As mentioned in previous sections, there are several prominent technological developments that need to occur before the realistic feasibility of the construction of the CCM, sometime in the next century, can be determined. Some of these developments will clearly occur, while others are of a more speculative nature. Below we list the major developments needed and classify them as either \emph{inevitable, highly likely, likely, speculative,} or \emph{highly speculative}.

\begin{itemize}
    
    \item \textbf{Magnet technology:} As discussed in Section~\ref{sec:time}, dipole magnets of 20~T can be taken for granted within the next few decades~\cite{botturasnowmass}. Thus, such a development, necessary for the CCM, can be classified as \emph{inevitable}.
    
    \item \textbf{Transportation:} As mentioned, there is strong interest from both public agencies and private companies in returning to the Moon, in order to establish bases or commercial operations, and there are multiple projects working toward this goal. It is possible that such projects will prove too difficult to make reliable and dependable, but there appears to be a strong enough interest to sustain such endeavors for the foreseeable future. Thus, reliable transportation between the Earth and the Moon in the next few decades seems \emph{highly likely}.
    
    \item \textbf{Robust Moon-to-Earth communication:} As discussed in Section~\ref{sec:other}, there are multiple advanced projects ongoing to facilitate and improve deep space communication. Thus, robust communication between the Moon and Earth, necessary for CCM operations and data transfer, seems \emph{highly likely}.

    \item \textbf{Lunar habitats, mining, and construction capabilities:} As discussed in Section~\ref{sec:time}, there are several projects underway that are pursuing clear goals for establishing habitats, conducting geological studies, and facilitating mining and construction operations on the Moon. Such projects are very ambitious and, as such, could be subject to delays. Because of this, the establishment, in the next century, of the lunar habitats, mining, and construction capabilities needed for the CCM is classified as \emph{likely}.
    
    \item \textbf{Lunar tunnel boring machines:} As discussed in Section~\ref{sec:time}, much work is currently ongoing in support of applying humanity's extensive experience with terrestrial tunnel boring machines to the geology of the Moon. Because such tunneling technology is well advanced, the eventual usage of LTBMs within the next 50 years seems \emph{likely}.
    
    \item \textbf{Remotely-operated lunar tunnel boring machines:} However, there are major challenges inherent to operating complex machines such as LTBMs remotely from the Earth, under the unusual conditions on the Moon (where unexpected difficulties seem inevitable). As such, the prospect of remotely-operated lunar tunnel boring machines in the next 50--100 years seems \emph{speculative}.
    
    \item \textbf{Lunar raw materials:} As discussed in Section~\ref{sec:dipoles}, initial studies of the lunar regolith suggest that sufficient raw materials --- particularly iron, silver, rare-earth elements, and pnictides --- could potentially be present. However, extensive studies can only be conducted once humans (or advanced surveying robots) return to the Moon. Thus, while the existence of raw materials needed to construct the HTS magnets of the CCM seems likely, it is more accurate to currently classify this prospect as \emph{speculative}.
    
    \item \textbf{Solar power on the Moon:} As discussed in Section~\ref{sec:powering}, the most likely candidate for a power source sufficient to power the CCM, on the level of 10~TW, is a solar panel Dyson belt around the equator of the Moon. The concept has been discussed for many decades and has been recently put on firmer conceptual and technological footing by the Shimizu Corporation, with their Luna Ring proposal~\cite{shimizu}, and also by JAXA, through  the development of the  
    Space Solar Power Systems~\cite{SSPS}, but much further work needs to be done. Given the highly ambitious nature of this proposal, the prospect of a Dyson belt around the equator of the Moon in the next 50 years to power (among other apparatuses) a CCM using 10~TW can be classified as \emph{highly speculative}. Because such a powering scheme is of the utmost importance for the CCM, and possibly for humankind in general, this prospect can also be determined as the highest priority for future research and development.
    
\end{itemize}

\newpage
\section{Conclusions}\label{sec:conclusions}

The construction of a very large hadron collider around the $\sim$11000-km circumference of a great circle of the Moon --- a Circular Collider on the Moon (CCM) --- is an attractive prospect for the (next-to-)next-to-next-generation of particle physics project, a potential successor to the proposed FCC at CERN, SPPC in China, or CitS in the Gulf of Mexico. The CCM could achieve an unprecedented center-of-mass proton-proton collision energy of 14 PeV, a thousand times more energetic than the LHC, utilizing dipole magnets that optimistically reach $\sim$20~T~\cite{botturasnowmass}. 

The CCM would require $\sim$5.5$ \times 10^{5}$ dipole magnets, which could be either ReBCO-based, requiring $\sim$7--13 k tons of rare-earth elements, or iron-based, requiring something on the order of a million tons of IBS, containing iron and pnictides. Many of the raw materials required to construct the machine, injector complex, detectors, and facilities of the CCM can potentially be sourced directly on the Moon.

Of the various scenarios for CCM construction discussed, Scenario C is the most compelling, which involves drilling shafts and excavating a nearly 11000-km tunnel --- along with additional tunneling for the injector complex and auxiliary tunnels --- a few dozen to a few hundred meters under the lunar surface to avoid lunary day-night temperature variations, cosmic radiation damage, and meteoroid strikes.

One plausible trajectory for the CCM would be to position it as circumnavigating the Moon approximately around the outer edge seen from the Earth, slightly displaced toward the Earth. This trajectory appears to avoid major elevation changes, and will be modified with more sophisticated tunnel-siting tools in the future.

Due to the extraordinary conditions of such a machine, including an energy loss per turn of $\sim$10$^{7}$ MeV and a synchrotron radiation power of 8.5 TW, a CCM powering system that relies upon abundant solar power will be needed. This could be achieved by constructing a belt of solar panels around the lunar equator at least 100 km in width, similar to what has been studied by, e.g., the Shimizu Corporation with the Luna Ring proposal.

The physics program of the CCM would be focused on exploring the unknown, searching for new physics beyond the Standard Model (BSM), as the scale of BSM physics is not concretely predictable and requires a purely exploratory mindset. The research at the CCM will be conducted by two detectors at the two interaction points, which could be built in a general-purpose manner, similar to ATLAS and CMS at the LHC. The detector design and technology required would be driven by the extreme conditions of the CCM such as an average number of proton interactions per bunch crossing on the level of $10^{6}$, orders of magnitude larger than the $\sim$200 at the HL-LHC.

A project such as the CCM will require global collaboration and could be one component of an entire Lunar Particle and Astrophysics Research Center (LPARC) hosting multiple experiments and projects.

Several large unknowns exist that will need to be addressed in detail before the ultimate feasibility of the CCM can be determined. The two most prominent among these are 1) whether the mineral composition of the lunar regolith contains sufficient raw materials (particular iron and rare-earth elements) to facilitate lunar-based construction of the components of the CCM; and 2) whether a large-scale solar power Dyson belt around the equator of the Moon can be constructed and operated realistically.

Under the assumption that the already excellent progress made in the development of industry-quality lunar tunnel boring machines, transport between Earth and Moon, detailed studies of lunar geology, solar powering on the Moon, and the construction of underground facilities for humans on the Moon continues apace during the next few decades, an 11000-km, 14-PeV Circular Collider on the Moon could, optimistically, start being constructed towards the end of the 21st century, and might then begin delivering data in the first half of the 22nd century. 

Such a project will be a very large step forward into the unknown, revealing  unparalleled information about the universe, as well as providing vital experience building large, non-terrestrial structures for humanity. The CCM will be an important stepping stone toward an ultimate Planck-scale collider, with a centre-of-mass energy of $\sim$10$^{16}$~TeV, that would require a minimum size equal to a tenth of the distance from Earth to Sun~\cite{Chen:1997ai,Zimmermann:2018koi}.

\newpage
\section{Acknowledgments}\label{sec:acknowledgments}
 
We thank Mark Robinson for helpful discussions and guidance about the surface and temperature profile of the Moon. We also thank Lucio Rossi and Davide Tommasini for enlightening discussions about high-temperature superconductors for use in accelerators. We are grateful to Qingjin Xu and Xinchou Lou for helpful information on the SPPC IBS magnet development program. With a considerable degree of  scepticism, Fritz Caspers has alerted us of the JAXA study. We further thank Jamal Rostami for illuminating discussions of the timeline for lunar tunnel boring machines.

This work was supported, in part, by the European Union's Horizon 2020 Research and Innovation Programme under grant agreement No.~101004730 (I.FAST). J.~Beacham acknowledges support from the U.S. Department of Energy's Office of Science under grant number DE-SC0010007.

\bibliography{ref}

\providecommand{\href}[2]{#2}\begingroup\raggedright\begin{thebibliography}{10}

\bibitem{Aad:2012tfa}
{\bfseries ATLAS} Collaboration, G.~Aad {\em et~al.}, ``{Observation of a new
  particle in the search for the Standard Model Higgs boson with the ATLAS
  detector at the LHC},''
  \href{http://dx.doi.org/10.1016/j.physletb.2012.08.020}{{\em Phys. Lett. B}
  {\bfseries 716} (2012) 1--29},
  \href{http://arxiv.org/abs/1207.7214}{{\ttfamily arXiv:1207.7214 [hep-ex]}}.

\bibitem{Chatrchyan:2012ufa}
{\bfseries CMS} Collaboration, S.~Chatrchyan {\em et~al.}, ``{Observation of a
  New Boson at a Mass of 125 GeV with the CMS Experiment at the LHC},''
  \href{http://dx.doi.org/10.1016/j.physletb.2012.08.021}{{\em Phys. Lett. B}
  {\bfseries 716} (2012) 30--61},
  \href{http://arxiv.org/abs/1207.7235}{{\ttfamily arXiv:1207.7235 [hep-ex]}}.

\bibitem{Abada:2019lih}
{\bfseries FCC} Collaboration, A.~Abada {\em et~al.}, ``{FCC Physics
  Opportunities}: {Future Circular Collider Conceptual Design Report Volume
  1},'' \href{http://dx.doi.org/10.1140/epjc/s10052-019-6904-3}{{\em Eur. Phys.
  J. C} {\bfseries 79} no.~6, (2019) 474}.

\bibitem{Abada:2019zxq}
{\bfseries FCC} Collaboration, A.~Abada {\em et~al.}, ``{FCC-ee: The Lepton
  Collider}: {Future Circular Collider Conceptual Design Report Volume 2},''
  \href{http://dx.doi.org/10.1140/epjst/e2019-900045-4}{{\em Eur. Phys. J. ST}
  {\bfseries 228} no.~2, (2019) 261--623}.

\bibitem{Benedikt:2018csr}
{\bfseries FCC} Collaboration, A.~Abada {\em et~al.}, ``{FCC-hh: The Hadron
  Collider}: {Future Circular Collider Conceptual Design Report Volume 3},''
  \href{http://dx.doi.org/10.1140/epjst/e2019-900087-0}{{\em Eur. Phys. J. ST}
  {\bfseries 228} no.~4, (2019) 755--1107}.

\bibitem{CEPC-SPPCStudyGroup:2015csa}
M.~Ahmad {\em et~al.}, ``{CEPC-SPPC Preliminary Conceptual Design Report. 1.
  Physics and Detector},''.

\bibitem{CEPC-SPPCStudyGroup:2015esa}
M.~Ahmad {\em et~al.}, ``{CEPC-SPPC Preliminary Conceptual Design Report. 2.
  Accelerator},''.

\bibitem{CEPCStudyGroup:2018rmc}
{\bfseries CEPC Study Group} Collaboration, ``{CEPC Conceptual Design Report:
  Volume 1 - Accelerator},'' \href{http://arxiv.org/abs/1809.00285}{{\ttfamily
  arXiv:1809.00285 [physics.acc-ph]}}.

\bibitem{CEPCStudyGroup:2018ghi}
{\bfseries CEPC Study Group} Collaboration, M.~Dong {\em et~al.}, ``{CEPC
  Conceptual Design Report: Volume 2 - Physics \& Detector},''
  \href{http://arxiv.org/abs/1811.10545}{{\ttfamily arXiv:1811.10545
  [hep-ex]}}.

\bibitem{McIntyre:2017ibd}
P.~M. McIntyre, S.~Assadi, S.~Bannert, J.~Breitschopf, D.~Chavez, J.~Gerity,
  J.~N. Kellams, N.~Pogue, and A.~Sattarov,
  \href{http://dx.doi.org/10.18429/JACoW-NAPAC2016-MOB2CO03}{``{Collider in the
  Sea: Vision for a 500 TeV World Laboratory},''} in {\em {2nd North American
  Particle Accelerator Conference}}, p.~MOB2CO03.
\newblock 2017.

\bibitem{Muong-2:2021vma}
{\bfseries Muon g-2} Collaboration, T.~Albahri {\em et~al.}, ``{Measurement of
  the anomalous precession frequency of the muon in the Fermilab Muon ``g-2"
  Experiment},'' \href{http://dx.doi.org/10.1103/PhysRevD.103.072002}{{\em
  Phys. Rev. D} {\bfseries 103} no.~7, (2021) 072002},
  \href{http://arxiv.org/abs/2104.03247}{{\ttfamily arXiv:2104.03247
  [hep-ex]}}.

\bibitem{2015hllh.book.....B}
O.~{Br{\"u}ning} and L.~{Rossi}, \href{http://dx.doi.org/10.1142/9581}{{\em
  {The High Luminosity Large Hadron Collider: The New Machine for Illuminating
  the Mysteries of Universe}}}.
\newblock 2015.

\bibitem{openstreetmap}
``{OpenStreetMap website}.'' \url{https://www.openstreetmap.org}.
\newblock Data is available under the Open Database License:
  https://www.openstreetmap.org/copyright; $\copyright$ OpenStreetMap
  contributors; Accessed: 2021-18-10.

\bibitem{wikimoon}
``{Full Moon photograph taken 10-22-2010 from Madison, Alabama, USA, from
  Wikipedia}.'' \url{https://en.wikipedia.org/wiki/File:FullMoon2010.jpg}.
\newblock Accessed: 2021-18-10.

\bibitem{siegrist}
{J.~Siegrist}, ``{NASA is going back to the moon. How could physicists
  collaborate?, in Remarks from Funding Agencies to the Snowmass Process:
  DOE}.'' {Snowmass Community Planning Meeting, 5 October 2020}, 2020.

\bibitem{artemis}
``{NASA Artemis Program}.''
\newblock \url{https://www.nasa.gov/artemisprogram}. Accessed: 2021-08-05.

\bibitem{CPLS}
``{NASA Commercial Lunar Payload Services}.''
\newblock \url{https://www.nasa.gov/content/commercial-lunar-payload-services}.
  Accessed: 2021-08-05.

\bibitem{doi:10.1061/0893}
F.~Ruess, J.~Schaenzlin, and H.~Benaroya, ``Structural design of a lunar
  habitat,''
  \href{http://dx.doi.org/10.1061/(ASCE)0893-1321(2006)19:3(133)}{{\em Journal
  of Aerospace Engineering} {\bfseries 19} no.~3, (2006) 133--157},
  \href{http://arxiv.org/abs/https://ascelibrary.org/doi/pdf/10.1061}{{\ttfamily
  https://ascelibrary.org/doi/pdf/10.1061}}.
  \url{https://ascelibrary.org/doi/abs/10.1061}.

\bibitem{ottr76}
{United Nations}, ``{United Nations Treaties and Principles on Outer Space},''
  {\em {ST/SPACE/11, ISBN 92-1-100900-6}} (Sep, 2002) .
  \url{http://www.unoosa.org/pdf/publications/STSPACE11E.pdf}.

\bibitem{botturasnowmass}
{L.~Bottura, et al.}, ``{A High Field Magnet Development for HEP in Europe –
  A Proposal},'' {\em {LOI submitted to Snowmass 2020}} (Sep, 2021) .
  \url{https://indico.cern.ch/event/999657/contributions/4207478/attachments/2179743/3681623/Snomass_HFM_LoI_v1.pdf}.

\bibitem{UGMM}
{Department of the Interior, U.S. Geological Survey}, ``{Unified Geologic Map
  of the Moon, 1:5M, 2020}.''
  \url{https://astrogeology.usgs.gov/search/map/Moon/Geology/Unified_Geologic_Map_of_the_Moon_GIS_v2},
  2020.
\newblock Accessed: 2021-21-03.

\bibitem{Cook:2141838}
C.~Cook, B.~Goddard, P.~Lebrun, J.~Osborne, Y.~Robert, C.~Sturzaker, M.~Sykes,
  Y.~Loo, J.~Brasser, and R.~Trunk, ``{Civil Engineering Optimisation Tool for
  the Study of CERN's Future Circular Colliders},''.
  \url{http://cds.cern.ch/record/2141838}.

\bibitem{nasamoontemp}
``{E}arth's {M}oon: {O}ur {N}atural {S}atellite, {N}{A}{S}{A} {S}cience
  website.'' \url{https://solarsystem.nasa.gov/moons/earths-moon/in-depth/}.
\newblock Accessed: 2021-16-03.

\bibitem{spey2016}
E.~Speyerer, R.~Povilaitis, M.~Robinson, {\em et~al.}, ``Quantifying crater
  production and regolith overturn on the moon with temporal imaging,''
  \href{http://dx.doi.org/https://doi.org/10.1038/nature19829}{{\em Nature}
  {\bfseries 538} (2016) 215--218}.
  \url{https://www.nature.com/articles/nature19829}.

\bibitem{nasachurnurl}
``{E}arth's {M}oon {H}it by {S}urprising {N}umber of {M}eteoroids, {N}{A}{S}{A}
  {G}oddard press release.''
  \url{https://www.nasa.gov/press-release/goddard/2016/lro-lunar-cratering},
  2016.
\newblock Accessed: 2021-16-03.

\bibitem{LCRTNASA}
``{Lunar Crater Radio Telescope (LCRT) on the Far-Side of the Moon, NASA
  website}.''
  \url{https://www.nasa.gov/directorates/spacetech/niac/2021_Phase_I/Lunar_Crater_Radio_Telescope/}.
\newblock Accessed: 2021-23-10.

\bibitem{1977NASSP.370..283L}
M.~B. {Langseth} and S.~J. {Keihm}, ``{In-situ measurements of lunar heat
  flow.},'' in {\em NASA Special Publication}, vol.~370, pp.~283--293.
\newblock 1977.

\bibitem{mendell1985}
W.~Mendell, ed., {\em Lunar Bases and Space Activities of the 21st Century}.
\newblock The Lunar and Planetary Institute, Houston, 1985.
\newblock \url{http://ads.harvard.edu/books/lbsa/}.

\bibitem{1988tamu.rept.....A}
C.~S. {Allen}, D.~W. {Cooper}, J.~{Davila}, David, C.~S. {Mahendra}, and M.~A.
  {Tagaras}, ``{Proposal for a lunar tunnel-boring machine}.'' {Final Report
  Texas A\&M Univ}, May, 1988.

\bibitem{rostami2018}
J.~Rostami, C.~Dreyer, and B.~Blair,
  \href{http://dx.doi.org/10.1061/9780784481899.024}{``{Lunar Tunnel Boring
  Machines},''} in {\em {Earth and Space 2018: Engineering for Extreme
  Environments}}, pp.~240 -- 252.
\newblock 2018.
\newblock \url{https://ascelibrary.org/doi/pdf/10.1061/9780784481899.024}.

\bibitem{tunneltalk}
``{T}{B}{M}s designed for excavations on the moon, {T}unnel{T}{E}{C}{H},
  {T}unnel {T}alk website.''
  \url{https://www.tunneltalk.com/TunnelTECH-Jun2019-Designing-TBMs-for-creating-infrastructure-on-the-moon.php},
  2019.
\newblock Accessed: 2021-18-03.

\bibitem{2017Icar..282...47B}
D.~M. {Blair}, L.~{Chappaz}, R.~{Sood}, C.~{Milbury}, A.~{Bobet}, H.~J.
  {Melosh}, K.~C. {Howell}, and A.~M. {Freed}, ``{The structural stability of
  lunar lava tubes},''
  \href{http://dx.doi.org/10.1016/j.icarus.2016.10.008}{{\em Icarus} {\bfseries
  282} (Jan., 2017) 47--55}.

\bibitem{melosh2018}
A.~{Theinat}, A.~{Modiriasari}, A.~{Bobet}, J.~{Melosh}, S.~{Dyke},
  J.~{Ramirez}, A.~{Maghareh}, and D.~{Gomez},
  \href{http://dx.doi.org/10.2514/6.2018-5185}{``Geometry and structural
  stability of lunar lava tubes,''} in {\em {2018 AIAA Space And Astronautics
  Forum And Exposition, Aiaa Space Forum, (aiaa 2018-5185)}}.
\newblock 2018.

\bibitem{garrick2014}
I.~Garrick-Bethell, V.~Perera, F.~Nimmo, and M.~Zuber, ``The tidal–rotational
  shape of the moon and evidence for polar wander,''
  \href{http://dx.doi.org/10.1038/nature13639}{{\em Nature} {\bfseries 512}
  (2014) 181–184}. \url{https://www.nature.com/articles/nature13639}.

\bibitem{lunarconst}
R.~B. Roncoli {\em et~al.}, ``Lunar constants and models.'' {J}{P}{L}
  {T}echnical {D}ocument {D}-32296, 2005.
\newblock \url{https://ssd.jpl.nasa.gov/?lunar_doc}. Accessed: 2021-18-03.

\bibitem{moonhigh}
M.~Robinson, ``Highest {P}oint on the {M}oon!.'' {L}unar {R}econnaissance
  {O}rbiter {C}amera website, {A}rizona {S}tate {U}niversity, 2010.
\newblock \url{http://lroc.sese.asu.edu/posts/249}. Accessed: 2021-18-03.

\bibitem{moonlow}
M.~Robinson, ``Great {W}all!.'' {L}unar {R}econnaissance {O}rbiter {C}amera
  website, {A}rizona {S}tate {U}niversity, 2016.
\newblock \url{http://lroc.sese.asu.edu/posts/898}. Accessed: 2021-18-03.

\bibitem{2010SSRv..150...81R}
M.~S. {Robinson}, S.~M. {Brylow}, M.~{Tschimmel}, D.~{Humm}, S.~J. {Lawrence},
  P.~C. {Thomas}, B.~W. {Denevi}, E.~{Bowman-Cisneros}, J.~{Zerr}, M.~A.
  {Ravine}, M.~A. {Caplinger}, F.~T. {Ghaemi}, J.~A. {Schaffner}, M.~C.
  {Malin}, P.~{Mahanti}, A.~{Bartels}, J.~{Anderson}, T.~N. {Tran}, E.~M.
  {Eliason}, A.~S. {McEwen}, E.~{Turtle}, B.~L. {Jolliff}, and H.~{Hiesinger},
  ``{Lunar Reconnaissance Orbiter Camera (LROC) Instrument Overview},''
  \href{http://dx.doi.org/10.1007/s11214-010-9634-2}{{\em Space Science
  Reviews} {\bfseries 150} no.~1-4, (Jan., 2010) 81--124}.

\bibitem{2016SSRv..200..431H}
D.~C. {Humm}, M.~{Tschimmel}, S.~M. {Brylow}, P.~{Mahanti}, T.~N. {Tran}, S.~E.
  {Braden}, S.~{Wiseman}, J.~{Danton}, E.~M. {Eliason}, and M.~S. {Robinson},
  ``{Flight Calibration of the LROC Narrow Angle Camera},''
  \href{http://dx.doi.org/10.1007/s11214-015-0201-8}{{\em Space Science
  Reviews} {\bfseries 200} no.~1-4, (Apr., 2016) 431--473}.

\bibitem{2016SSRv..200..393M}
P.~{Mahanti}, D.~C. {Humm}, M.~S. {Robinson}, A.~K. {Boyd}, R.~{Stelling},
  H.~{Sato}, B.~W. {Denevi}, S.~E. {Braden}, E.~{Bowman-Cisneros}, S.~M.
  {Brylow}, and M.~{Tschimmel}, ``{Inflight Calibration of the Lunar
  Reconnaissance Orbiter Camera Wide Angle Camera},''
  \href{http://dx.doi.org/10.1007/s11214-015-0197-0}{{\em Space Science
  Reviews} {\bfseries 200} no.~1-4, (Apr., 2016) 393--430}.

\bibitem{2012ISPAr39B4..511S}
E.~J. {Speyerer}, R.~V. {Wagner}, M.~S. {Robinson}, D.~C. {Humm}, K.~{Becker},
  J.~{Anderson}, and P.~{Thomas}, ``{In-Flight Geometric Calibration of the
  Lunar Reconnaissance Orbiter Camera},''
  \href{http://dx.doi.org/10.5194/isprsarchives-XXXIX-B4-511-2012}{{\em ISPRS -
  International Archives of the Photogrammetry, Remote Sensing and Spatial
  Information Sciences} {\bfseries 39B4} (Aug., 2012) 511--516}.

\bibitem{LROC:EDR}
M.~S. Robinson, ``{Lunar Reconnaissance Orbiter Camera Experimental Data
  Record, LRO-L-LROC-2-EDR-V1.0, NASA Planetary Data System},''.
  \url{https://pds.nasa.gov/ds-view/pds/viewDataset.jsp?dsid=LRO-L-LROC-2-EDR-V1.0}.

\bibitem{2011JE003926}
F.~Scholten, J.~Oberst, K.-D. Matz, T.~Roatsch, M.~Wählisch, E.~J. Speyerer,
  and M.~S. Robinson, ``{GLD100: The near-global lunar 100 m raster DTM from
  LROC WAC stereo image data},''
  \href{http://dx.doi.org/https://doi.org/10.1029/2011JE003926}{{\em Journal of
  Geophysical Research: Planets} {\bfseries 117} no.~E12, (2012) }.
  \url{https://agupubs.onlinelibrary.wiley.com/doi/abs/10.1029/2011JE003926}.

\bibitem{2010GL043751}
D.~E. Smith, M.~T. Zuber, G.~A. Neumann, F.~G. Lemoine, E.~Mazarico, M.~H.
  Torrence, J.~F. McGarry, D.~D. Rowlands, J.~W. Head~III, T.~H. Duxbury,
  O.~Aharonson, P.~G. Lucey, M.~S. Robinson, O.~S. Barnouin, J.~F. Cavanaugh,
  X.~Sun, P.~Liiva, D.-d. Mao, J.~C. Smith, and A.~E. Bartels, ``{Initial
  observations from the Lunar Orbiter Laser Altimeter (LOLA)},''
  \href{http://dx.doi.org/https://doi.org/10.1029/2010GL043751}{{\em
  Geophysical Research Letters} {\bfseries 37} no.~18, (2010) }.
  \url{https://agupubs.onlinelibrary.wiley.com/doi/abs/10.1029/2010GL043751}.

\bibitem{marehumboldt}
``{H}umboldtianum {B}asin, {N}{A}{S}{A} {L}unar {R}econnaissance {O}rbiter
  website.''
  \url{https://www.nasa.gov/mission_pages/LRO/multimedia/lroimages/lola-20100604_humboldt.html},
  2010.
\newblock Accessed: 2021-18-03.

\bibitem{PhysRevSTAB.18.101002}
M.~Benedikt, D.~Schulte, and F.~Zimmermann, ``Optimizing integrated luminosity
  of future hadron colliders,''
  \href{http://dx.doi.org/10.1103/PhysRevSTAB.18.101002}{{\em Phys. Rev. ST
  Accel. Beams} {\bfseries 18} (Oct, 2015) 101002}.
  \url{https://link.aps.org/doi/10.1103/PhysRevSTAB.18.101002}.

\bibitem{Assmann:453821}
R.~W. Assmann and K.~Cornelis, ``{The Beam-Beam Interaction in the Presence of
  Strong Radiation Damping},''. \url{https://cds.cern.ch/record/453821}.

\bibitem{bordry2018machine}
F.~Bordry, M.~Benedikt, O.~Bruning, J.~Jowett, L.~Rossi, D.~Schulte,
  S.~Stapnes, and F.~Zimmermann, ``{Machine Parameters and Projected Luminosity
  Performance of Proposed Future Colliders at CERN},'' {\em arXiv: 1810.13022}
  (2018) .

\bibitem{Menon:2012dr}
M.~J. Menon and P.~V. R.~G. Silva, ``{An updated analysis on the rise of the
  hadronic total cross-section at the LHC energy region},''
  \href{http://dx.doi.org/10.1142/S0217751X13500991}{{\em Int. J. Mod. Phys. A}
  {\bfseries 28} (2013) 1350099},
  \href{http://arxiv.org/abs/1212.5096}{{\ttfamily arXiv:1212.5096 [hep-ph]}}.

\bibitem{MFS}
``{NASA Moon Fact Sheet}.''
  \url{https://nssdc.gsfc.nasa.gov/planetary/factsheet/moonfact.html}, 2020.
\newblock Accessed: 2021-25-03.

\bibitem{Bruning:782076}
O.~S. Brüning, P.~Collier, P.~Lebrun, S.~Myers, R.~Ostojic, J.~Poole, and
  P.~Proudlock, \href{http://dx.doi.org/10.5170/CERN-2004-003-V-1}{{\em {LHC
  Design Report}}}.
\newblock CERN Yellow Reports: Monographs. CERN, Geneva, 2004.
\newblock \url{http://cds.cern.ch/record/782076}.

\bibitem{globalpower}
``{Energy Production and Consumption, Our World in Data website}.''
  \url{https://ourworldindata.org/energy-production-consumption}.
\newblock Accessed: 2021-15-05.

\bibitem{CLIMENT2014352}
B.~Climent, O.~Torroba, R.~González-Cinca, N.~Ramachandran, and M.~D. Griffin,
  ``Heat storage and electricity generation in the moon during the lunar
  night,''
  \href{http://dx.doi.org/https://doi.org/10.1016/j.actaastro.2013.07.024}{{\em
  Acta Astronautica} {\bfseries 93} (2014) 352--358}.
  \url{https://www.sciencedirect.com/science/article/pii/S0094576513002567}.

\bibitem{nasafission}
``{NASA Developing Fission Surface Power Technology press release website}.''
  \url{https://www.nasa.gov/home/hqnews/2008/sep/HQ_08-227_Moon_Power.html},
  2008.
\newblock Accessed: 2021-11-04.

\bibitem{doefission1}
``{U.S. DoE, Office of Nuclear Energy, 5 Things You Need to Know about Fission
  Surface Power Systems website}.''
  \url{https://www.energy.gov/ne/articles/5-things-you-need-know-about-fission-surface-power-systems},
  2020.
\newblock Accessed: 2021-11-04.

\bibitem{doefission2}
``{U.S. General Services Administration, Request for Information for Fission
  Surface Power (FSP) website}.''
  \url{https://beta.sam.gov/opp/b92644af4767413c9e831e633c6e2888/view}, 2020.
\newblock Accessed: 2021-11-04.

\bibitem{mitnuclear}
``{A} {L}unar {N}uclear {R}eactor, {M}{I}{T} {T}echnology {R}eview website.''
  \url{https://www.technologyreview.com/2009/08/17/210973/a-lunar-nuclear-reactor/},
  2009.
\newblock Accessed: 2021-16-03.

\bibitem{iaea}
``{IAEA PRIS: Operational \& Long-Term Shutdown Reactors website}.''
  \url{https://pris.iaea.org/PRIS/WorldStatistics/OperationalReactorsByCountry.aspx},
  2021.
\newblock Accessed: 2021-11-04.

\bibitem{iter}
{ITER Organization}, ``{I}{T}{E}{R} {R}esearch {P}lan within the {S}taged
  {A}pproach ({L}evel {I}{I}{I} - {P}rovisional {V}ersion)
  ({I}{T}{R}--18-003),'' 2018.
\newblock
  \url{https://inis.iaea.org/search/searchsinglerecord.aspx?recordsFor=SingleRecord&RN=50056648}.

\bibitem{1367615}
A.~Ignatiev, A.~Freundlich, and C.~Horton,
  \href{http://dx.doi.org/10.1109/AERO.2004.1367615}{``Solar cell development
  on the surface of moon from in-situ lunar resources,''} in {\em 2004 IEEE
  Aerospace Conference Proceedings (IEEE Cat. No.04TH8720)}, vol.~1, p.~318
  Vol.1.
\newblock 2004.

\bibitem{Dyson1667}
F.~J. Dyson, ``{Search for Artificial Stellar Sources of Infrared Radiation},''
  \href{http://dx.doi.org/10.1126/science.131.3414.1667}{{\em Science}
  {\bfseries 131} no.~3414, (1960) 1667--1668}.
  \url{https://science.sciencemag.org/content/131/3414/1667}.

\bibitem{NSFS}
``{NASA Sun Fact Sheet}.''
  \url{https://nssdc.gsfc.nasa.gov/planetary/factsheet/sunfact.html}, 2020.
\newblock Accessed: 2021-23-03.

\bibitem{nasasolar}
``{NASA: Solar Power Investigation to Launch on Lunar Lander website}.''
  \url{https://www.nasa.gov/feature/glenn/2020/solar-power-investigation-to-launch-on-lunar-lander},
  2020.
\newblock Accessed: 2021-11-04.

\bibitem{shimizu}
``{Shimizu: LUNA RING, Solar Power Generation on the Moon website}.''
  \url{https://www.shimz.co.jp/en/topics/dream/content02/ }.
\newblock Accessed: 2021-11-04.

\bibitem{criswellLSP}
D.~Criswell, ``Lunar solar power,''
  \href{http://dx.doi.org/10.1109/MP.2004.1301242}{{\em Potentials, IEEE}
  {\bfseries 22} (02, 2004) 20 -- 25}.

\bibitem{SSPS}
``{JAXA: Research on the Space Solar Power Systems (SSPS)}.''
  \url{https://www.kenkai.jaxa.jp/eng/research/ssps/ssps-index.html/ }.
\newblock Accessed: 2021-04-05.

\bibitem{muoncollider2020}
N.~Bartosik, A.~Bertolin, L.~Buonincontri, M.~Casarsa, F.~Collamati,
  A.~Ferrari, A.~Ferrari, A.~Gianelle, D.~Lucchesi, N.~Mokhov, M.~Palmer,
  N.~Pastrone, P.~Sala, L.~Sestini, and S.~Striganov, ``Detector and physics
  performance at a muon collider,''.
  \url{https://doi.org/10.1088/1748-0221/15/05/p05001}.

\bibitem{King:1999gb}
B.~J. King, ``{Potential hazards from neutrino radiation at muon colliders},''
  \href{http://arxiv.org/abs/physics/9908017}{{\ttfamily
  arXiv:physics/9908017}}.

\bibitem{vanNugteren:2018xal}
J.~van Nugteren, G.~Kirby, J.~Murtom\"aki, G.~De~Rijk, L.~Rossi, and
  A.~Stenvall, ``{Towards REBCO 20T+ dipoles for accelerators},''
  \href{http://dx.doi.org/10.1109/TASC.2018.2820177}{{\em IEEE Trans. Appl.
  Supercond.} {\bfseries 28} no.~4, (2018) 4008509}.

\bibitem{Bonura_2014}
M.~Bonura and C.~Senatore, ``High-field thermal transport properties of rebco
  coated conductors,''
  \href{http://dx.doi.org/10.1088/0953-2048/28/2/025001}{{\em Superconductor
  Science and Technology} {\bfseries 28} no.~2, (Dec, 2014) 025001}.
  \url{http://dx.doi.org/10.1088/0953-2048/28/2/025001}.

\bibitem{Assmann:1284326}
R.~Assmann, R.~Bailey, O.~Brüning, O.~Dominguez~Sanchez, G.~de~Rijk, J.~M.
  Jimenez, S.~Myers, L.~Rossi, L.~Tavian, E.~Todesco, and F.~Zimmermann,
  ``{First Thoughts on a Higher-Energy LHC},'' tech. rep., CERN, Geneva, Aug,
  2010.
\newblock \url{http://cds.cern.ch/record/1284326}.

\bibitem{globalREE}
``{Reserves of rare earths worldwide from 2010 to 2020, Statista website}.''
  \url{https://www.statista.com/statistics/271874/rare-earths-global-reserves/}.
\newblock Accessed: 2021-06-04.

\bibitem{resources6030040}
C.~L. McLeod and M.~P.~S. Krekeler, ``{Sources of Extraterrestrial Rare Earth
  Elements: To the Moon and Beyond},''
  \href{http://dx.doi.org/10.3390/resources6030040}{{\em Resources} {\bfseries
  6} no.~3, (2017) }. \url{https://www.mdpi.com/2079-9276/6/3/40}.

\bibitem{Schultz468}
P.~H. Schultz, B.~Hermalyn, A.~Colaprete, K.~Ennico, M.~Shirley, and W.~S.
  Marshall, ``{The LCROSS Cratering Experiment},''
  \href{http://dx.doi.org/10.1126/science.1187454}{{\em Science} {\bfseries
  330} no.~6003, (2010) 468--472}.
  \url{https://science.sciencemag.org/content/330/6003/468}.

\bibitem{ironhts}
Y.~Kamihara, T.~Watanabe, M.~Hirano, and H.~Hosono, ``{Iron-based layered
  superconductor La[O1-xFx]FeAs (x= 0.05-0.12) with Tc = 26 K},''
  \href{http://dx.doi.org/10.1021/ja800073m}{{\em Journal of the American
  Chemical Society} {\bfseries 130} no.~11, (Mar., 2008) 3296--3297}.

\bibitem{crawford2014lunar}
I.~A. Crawford, ``{Lunar Resources: A Review},''
  \href{http://arxiv.org/abs/1410.6865}{{\ttfamily arXiv:1410.6865
  [astro-ph.EP]}}.

\bibitem{HEGGY2020116274}
E.~Heggy, E.~Palmer, T.~Thompson, B.~Thomson, and G.~Patterson, ``{Bulk
  composition of regolith fines on lunar crater floors: Initial investigation
  by LRO/Mini-RF},''
  \href{http://dx.doi.org/https://doi.org/10.1016/j.epsl.2020.116274}{{\em
  Earth and Planetary Science Letters} {\bfseries 541} (2020) 116274}.
  \url{https://www.sciencedirect.com/science/article/pii/S0012821X2030217X}.

\bibitem{Zhang_2021}
Z.~Zhang, D.~Wang, S.~Wei, Y.~Wang, C.~Wang, Z.~Zhang, H.~Yao, X.~Zhang,
  F.~Liu, H.~Liu, Y.~Ma, Q.~Xu, and Y.~Wang, ``{First performance test of the
  iron-based superconducting racetrack coils at 10 T},''
  \href{http://dx.doi.org/10.1088/1361-6668/abb11b}{{\em Superconductor Science
  and Technology} {\bfseries 34} no.~3, (Feb, 2021) 035021}.
  \url{https://doi.org/10.1088/1361-6668/abb11b}.

\bibitem{Wang_2019}
D.~Wang, Z.~Zhang, X.~Zhang, D.~Jiang, C.~Dong, H.~Huang, W.~Chen, Q.~Xu, and
  Y.~Ma, ``{First performance test of a 30 mm iron-based superconductor single
  pancake coil under a 24 T background field},''
  \href{http://dx.doi.org/10.1088/1361-6668/ab09a4}{{\em Superconductor Science
  and Technology} {\bfseries 32} no.~4, (Mar, 2019) 04LT01}.
  \url{https://doi.org/10.1088/1361-6668/ab09a4}.

\bibitem{doi:10.1142/S0217751X19400037}
E.~Kong, C.~Wang, L.~Wang, X.~Wang, D.~Cheng, K.~Zhang, Y.~Wang, Q.~Peng, and
  Q.~Xu, ``{Conceptual design study of iron-based superconducting dipole
  magnets for SPPC},'' \href{http://dx.doi.org/10.1142/S0217751X19400037}{{\em
  International Journal of Modern Physics A} {\bfseries 34} no.~13n14, (2019)
  1940003},
  \href{http://arxiv.org/abs/https://doi.org/10.1142/S0217751X19400037}{{\ttfamily
  https://doi.org/10.1142/S0217751X19400037}}.
  \url{https://doi.org/10.1142/S0217751X19400037}.

\bibitem{Tang2017}
{J.~Tang}, ``{SPPC Study Progress},'' {\em {FCC Week 2017, Berlin}} .
  \url{https://indico.cern.ch/event/556692}.

\bibitem{QingjinXu}
{Qingjin Xu}, ``{Private communication},'' 2021.

\bibitem{rostamipriv}
{Jamal Rostami}, ``{Private communication},'' 2021.

\bibitem{spectrumilrs}
{Andrew Jones}, ``{China Aims for a Permanent Moon Base in the 2030s. Lunar
  megaproject to be a stepping-stone to the solar system},'' {\em {IEEE
  Spectrum}} ({Sep}, 2021) .
  \url{https://spectrum.ieee.org/china-aims-for-a-permanent-moon-base-in-the-2030s}.

\bibitem{ilrsspace}
{Tereza Pultarova}, ``{Russia, China reveal moon base roadmap but no plans for
  astronaut trips yet},'' {\em {Space.com}} ({Jun}, 2021) .
  \url{https://www.space.com/china-russia-international-lunar-research-station}.

\bibitem{7119255}
D.~W. Scott, P.~A. Curreri, C.~K. Ferguson, M.~E. Nall, M.~L. Tinker, and G.~M.
  Wright, \href{http://dx.doi.org/10.1109/AERO.2015.7119255}{``Germinating the
  2050 cis-lunar econosphere,''} in {\em 2015 IEEE Aerospace Conference},
  pp.~1--16.
\newblock 2015.

\bibitem{usaf}
{Air Force Space Command, Peterson AFB, United States}, ``{The Future of Space
  2060 and Implications for U.S. Strategy: Report on the Space Futures
  Workshop},'' {\em {Technical Report AD1095527}} ({Mar and Sep}, 2019) .
  \url{https://apps.dtic.mil/sti/pdfs/AD1095527.pdf}.

\bibitem{wlcg}
``{Worldwide LHC Computing Grid, CERN website}.''
  \url{https://wlcg-public.web.cern.ch/}.
\newblock Accessed: 2021-23-10.

\bibitem{NASASCAN}
``{Space Communications and Navigation, NASA website}.''
  \url{https://www.nasa.gov/directorates/heo/scan/index.html}.
\newblock Accessed: 2021-23-10.

\bibitem{ESAcomm}
``{Deep space communication and navigation, ESA website}.''
  \url{https://www.esa.int/Enabling_Support/Preparing_for_the_Future/Discovery_and_Preparation/Deep_space_communication_and_navigation}.
\newblock Accessed: 2021-23-10.

\bibitem{JAXAcomm}
``{Low-cost, High-capacity, High-speed Satellite Communication System for
  Society5.0, JAXA website}.''
  \url{https://www.kenkai.jaxa.jp/eng/research/society5/society5.html}.
\newblock Accessed: 2021-23-10.

\bibitem{lhcions}
{Bruce, R. and d'Enterria, D. and de Roeck, A. and others}, ``{New physics
  searches with heavy-ion collisions at the LHC},'' {\em J. Phys. G} {\bfseries
  47} (Dec, 2018) 060501. 20 p,
  \href{http://arxiv.org/abs/1812.07688}{{\ttfamily arXiv:1812.07688}}.
  \url{http://cds.cern.ch/record/2652738}.

\bibitem{fcchhions}
{Dainese, A. and Wiedemann, U.A. and Armesto, N. and others}, ``{Heavy ions at
  the Future Circular Collider. Heavy ions at the Future Circular Collider},''
  \href{http://dx.doi.org/10.23731/CYRM-2017-003.635}{{\em CERN Yellow Report}
  (May, 2016) 635--692. 58 p},
  \href{http://arxiv.org/abs/1605.01389}{{\ttfamily arXiv:1605.01389}}.
  \url{http://cds.cern.ch/record/2150910}.

\bibitem{1992LPI....23.1537W}
T.~L. {Wilson}, ``{An Earth-Moon Baseline Neutrino Experiment},'' in {\em Lunar
  and Planetary Science Conference}, vol.~23 of {\em Lunar and Planetary
  Science Conference}, p.~1537.
\newblock Mar., 1992.

\bibitem{Chen:1997ai}
P.~Chen and R.~J. Noble, ``{Crystal channel collider: Ultrahigh-energy and
  luminosity in the next century},''
  \href{http://dx.doi.org/10.1063/1.53055}{{\em AIP Conf. Proc.} {\bfseries
  398} no.~1, (1997) 273--285}.

\bibitem{Zimmermann:2018koi}
F.~Zimmermann, ``{Future colliders for particle physics
  \textemdash{}\textquotedblleft{}Big and small\textquotedblright{}},''
  \href{http://dx.doi.org/10.1016/j.nima.2018.01.034}{{\em Nucl. Instrum. Meth.
  A} {\bfseries 909} (2018) 33--37},
  \href{http://arxiv.org/abs/1801.03170}{{\ttfamily arXiv:1801.03170
  [physics.acc-ph]}}.

\end{thebibliography}\endgroup
\bibliographystyle{utphys}

\bibliographystyle{utphys}

\end{document}